\documentclass[pra,onecolumn,superscriptaddress,amsmath,amssymb]{revtex4}

\usepackage{graphicx}
\usepackage{dcolumn}
\usepackage{bm}
\usepackage{hyperref}
\usepackage{amsmath}
\usepackage{amssymb}
\usepackage{subfigure}
\usepackage{epstopdf}
\usepackage{color}
\usepackage{wasysym}

\begin{document}

\begin{abstract}
We consider a correlated Bose gas tightly confined into a ring shaped lattice,
  in the presence of an artificial gauge potential inducing a persistent current through it.
  A weak link painted on the ring acts as a source of coherent back-scattering
  for the propagating gas, interfering with the forward scattered current.
  This system defines an atomic counterpart of the rf-SQUID:
  the atomtronics quantum interference device (AQUID).
  The goal of the present study is to corroborate the emergence of an effective two-level system
  in such a setup and to assess its quality, in terms of its inner resolution
  and its separation from the rest of the many-body spectrum,
  across the different physical regimes.
  In order to achieve this aim, we examine the dependence of the qubit energy gap	
  on the bosonic density, the interaction strength, and the barrier depth,
  and we show how the superposition between current states appears
  in the momentum distribution (time-of-flight) images.
  A mesoscopic ring lattice with intermediate-to-strong interactions and weak barrier depth
  is found to be a favorable candidate for setting up,
  manipulating and probing a qubit in the next generation of atomic experiments.
\end{abstract}

\title{Coherent superposition of current flows in an Atomtronic Quantum Interference Device}

\author{Davit Aghamalyan}
\affiliation{Centre for Quantum Technologies, National University of Singapore, Singapore 117543}

\author{Marco Cominotti}
\affiliation{Universit\'e Grenoble Alpes, LPMMC, F-38000 Grenoble, France}
\affiliation{CNRS, LPMMC, F-38000 Grenoble, France}

\author{Matteo Rizzi}
\affiliation{Institut f\"ur Physik, Johannes Gutenberg-Universit\"at Mainz,Staudingerweg 7, D-55099 Mainz, Germany}

\author{Davide Rossini}
\affiliation{NEST, Scuola Normale Superiore and Istituto Nanoscienze-CNR, I-56126 Pisa, Italy}

\author{Frank Hekking}
\affiliation{Universit\'e Grenoble Alpes, LPMMC, F-38000 Grenoble, France}
\affiliation{CNRS, LPMMC, F-38000 Grenoble, France}

\author{Anna Minguzzi}
\affiliation{Universit\'e Grenoble Alpes, LPMMC, F-38000 Grenoble, France}
\affiliation{CNRS, LPMMC, F-38000 Grenoble, France}

\author{Leong-Chuan Kwek}
\affiliation{Centre for Quantum Technologies, National University of Singapore, Singapore 117543}
\affiliation{National Institute of Education and Institute of Advanced Studies, Nanyang Technological University, Singapore 637616}

\author{Luigi Amico}
\affiliation{CNR-MATIS-IMM $\&$ Dipartimento di Fisica e Astronomia,  I-95127 Catania, Italy}
\affiliation{Centre for Quantum Technologies, National University of Singapore, Singapore 117543}

\maketitle


\section{Introduction}
The progress achieved in optical micro-fabrication has led to the foundation of {\it atomtronics}:
Bose-Einstein condensates manipulated with lithographic precision in optical circuits of various intensities and 
spatial shapes~\cite{Seaman,Birkl,litho_Dumke,litho_Schlosser,Boshier_bessel}. 
The neutrality of the atoms carrying the current (substantially reducing decoherence sources),
the flexibility on their statistics (bosonic/fermionic) and interactions (tunable from short
to long-range, from attractive to repulsive) are some of the key features of atomtronic circuits.
Atomtronics sets a new stage for quantum simulations~\cite{coldatoms},
with remarkable spin-offs in other fields of science and technology.
This activity is believed to lead, in turn, to an improved understanding of actual electronic systems.

An important representative example of an atomtronic circuit is provided
by a Bose-Einstein condensate flowing in a ring-shaped trapping
potential~\cite{ring_Franke,ring_Henderson,AQUID_phillips,ring_Morizot,ring_dressed_Foot,foot,nigel, luigi,turpin}.
A barrier potential painted along the ring originates a weak link, acting as a source
of back-scattering for the propagating condensate,
thus creating an interference state with the forward scattered current.
This gives rise to an atomic condensate counterpart of the celebrated
rf-SQUID---a superconducting ring interrupted by a Josephson junction~\cite{flux-qubit,Tinkham},
namely an Atomtronic Quantum Interference Device (AQUID).
Due to the promising combination of advantages characterizing Josephson junctions and cold atoms,
the AQUID has been the object of recent investigations~\cite{AQUID_Eckel,AQUID_resistive}.
The first experimental realizations made use of a Bose-Einstein condensate
free to move along a toroidal potential, except through a small spatial region,
where an effective potential constriction (giving rise to the aforementioned weak link) is created via a very focused blue-detuned laser or via a painting potential~\cite{ramanathan,AQUID_Boshier,wright,Eckel}.

On the theoretical side, it has been demonstrated that the two currents flowing
in the AQUID can, indeed, define an effective two-level system,
that is, the cold-atom analog of flux qubits~\cite{solenov_qubit,Amico_qubit}.
The potential constriction breaks the Galilean invariance and splits the qubit levels
that otherwise would be perfectly degenerate at half-flux quantum.
In this context, it is of vital importance for the qubit dynamics
that a good energy resolution of the two levels could be achieved in realistic physical situations
(while keeping the qubit well separated from the rest of the many-body spectrum).


In this paper we focus on ring-shaped confinements with a {\it lattice modulation}
and a {\it potential constriction}. This set-up, that can be realized following different
routes (see, e.g., Ref.~\cite{Amico_qubit}), presents several advantages for the design of an AQUID.
First of all, assuming that the bosons occupy only the lowest Bloch band~\footnote{ This condition is especially feasible nowadays, because the gap between the lowest Bloch bands can be magnified, by playing with the shape of the wells, a feature that is straightforward to implement realizing the ring lattice with SLM devices. The influence of the other Bloch bands has been analyzed in~\cite{buchler}. },
the ring lattice helps in controlling the current. For instance, because of the one-dimensional dynamics, the vortex formation rate along the flow is negligible.
Secondly, it helps localizing the barrier effect to a point-like scale with respect
to lattice spacing, which should in turn yield a favorable scaling of the qubit gap
with the bosonic density~\cite{rey_non-homo}.
Moreover, it provides an easy route to realize interacting ring-ring
architectures~\cite{luigi,Amico_qubit}~\footnote{Experimentally, the ring lattice are arranged along a laser confinement with cylindrical symmetry, with a `pancake' structure. The the inter-ring tunneling, however can be made negligible with different approaches (for example suitably focusing the laser beams). See also Ref.~\cite{couple_rings}.}.

This issue has been considered so far only in some limiting cases, e.g. for particular types of superposition states or in the infinitely strong interacting regime~\cite{hallwood_homogeneous,rey_non-homo}.
We perform a systematic study  on the quality of the qubit in the cold-atom ring lattice:
in particular, we characterize the energy structure at the degeneracy point
at half-flux quantum, and study how it is possible to observe experimentally the superposition of the current flows. By employing a combination of analytical and numerical techniques,
that allows us to cover all the relevant physical regimes of system sizes, filling, barrier and interaction strengths, we show that:
{\it i)} the gap $\Delta E_1$
between the states of the effective two-level system scales as a power law with the system size;
{\it ii)} at a mesoscopic scale, a qubit is well-defined, with $\Delta E_1$ displaying
a favorable dependence in a wide range of system parameters;
{\it iii)} the superposition state is {detectable in the momentum distribution of the bosonic gas, which is
measurable via time-of-flight (TOF) expansion,
and {\it iv)} the momentum distribution exhibits a subtle interplay
between barrier strength and interaction.

The paper is organized as follows. In the next section we present
the physical system of interacting bosons on a 1D ring lattice with a potential constriction,
and the effective two-level system giving rise to the AQUID.
In Section~\ref{Sect:gaps} the energy spectrum of the system and its scaling with system size,
filling, and interaction is analyzed.
In Section~\ref{Sect:Momentum} we show how the state of the AQUID can be read out
through TOF expansion images of the gas. 
Finally, we draw our conclusions in Section~\ref{RingsVsContinuous}.
Technicalities on the employed methods and further details are provided in the Appendices.

\section{The physical system}
\label{Sect:model}

We consider a system of $N$ interacting bosons at zero temperature,
loaded into a 1D ring-shaped optical lattice of $M$ sites.
The discrete rotational symmetry of the lattice ring is broken by the presence
of a localized potential on one lattice site, which gives rise to a weak link.
The ring is pierced by an artificial (dimensionless) magnetic flux $\Omega$,
which can be experimentally induced for neutral atoms as a Coriolis flux by rotating the lattice
at constant velocity~\cite{fetter,wright}, or as a synthetic gauge flux by imparting a geometric phase
directly to the atoms via suitably designed laser fields~\cite{berry,synth1,dalibard}.

In the tight-binding approximation, this system is described
by the 1D Bose-Hubbard (BH) Hamiltonian
\begin{equation}
  H \! = \! \sum_{j=1}^{M} \left[ \! -t (e^{-i\Omega/M}b_{j}^{\dagger}b_{j+1} \! + \! {\rm h.c.} \!)
  + \! \frac{U}{2} n_{j}(n_{j} \! - \! 1) \! + \! \Lambda_{j} n_{j} \! \right],
  \label{Model}
\end{equation}
where $b_j \, (b^{\dagger}_j)$ are bosonic annihilation (creation) operators on the $j$th site
and $n_{j}= b_{j}^{\dag}b_{j}$ is the corresponding number operator.
Periodic boundaries are imposed by taking $b_{M+1} \equiv b_1$.
The parameter $U$ takes into account the finite scattering length for the atomic 
two-body collisions on the same site; $\Lambda_j$ defines an externally applied local potential.
Periodic boundary conditions are assumed in order to account for the multiply connected geometry of the ring system.
The presence of the flux $\Omega$ is taken into account through the Peierls substitution:
$t\rightarrow t e^{-i\Omega/M}$ ($t$ is the hopping amplitude).
In the thermodynamic limit, the BH model for $\Lambda_{j}=0 \;\forall j$ displays 
a superfluid to Mott-insulator transition for commensurate fillings $N/M$, 
and at a critical value of the ratio $U/t$ of interaction-to-tunnel energy.
The phase boundaries of the transition are expected to be affected by the magnetic flux
through an overall rescaling $t / U \rightarrow (t / U) \cos(\Omega/M)$~\cite{monien}.
The potential barrier considered here is localized on a single site $j_0$,
i.e., $\Lambda_{j}=\Lambda\delta_{j,j_0}$ with $\delta_{i,j}$ being the Kronecker delta.
As we will discuss in Section~\ref{Sect:gap_mesoscopic}, we find a superfluid-insulator transition 
even if the ring is interrupted by a weak link, although the phenomenon 
is rather a crossover, the ring being of finite size. 

In this work, specific regimes of the system described by Eq.~\eqref{Model} will be captured
analytically via the Tonks-Girardeau (TG) mapping (hard-core limit of infinite repulsions),
and the mean-field Gross-Pitaevskii (GP) approximation (weak interactions and large fillings).
To cover all the interaction regimes, numerical analysis will be also pursued, through truncated
and exact diagonalization (ED) schemes and density-matrix renormalization-group (DMRG) methods.
Details on these techniques are given in~\ref{methods}.

\begin{figure}[!t]
\centering
  \includegraphics[width=0.5\textwidth]{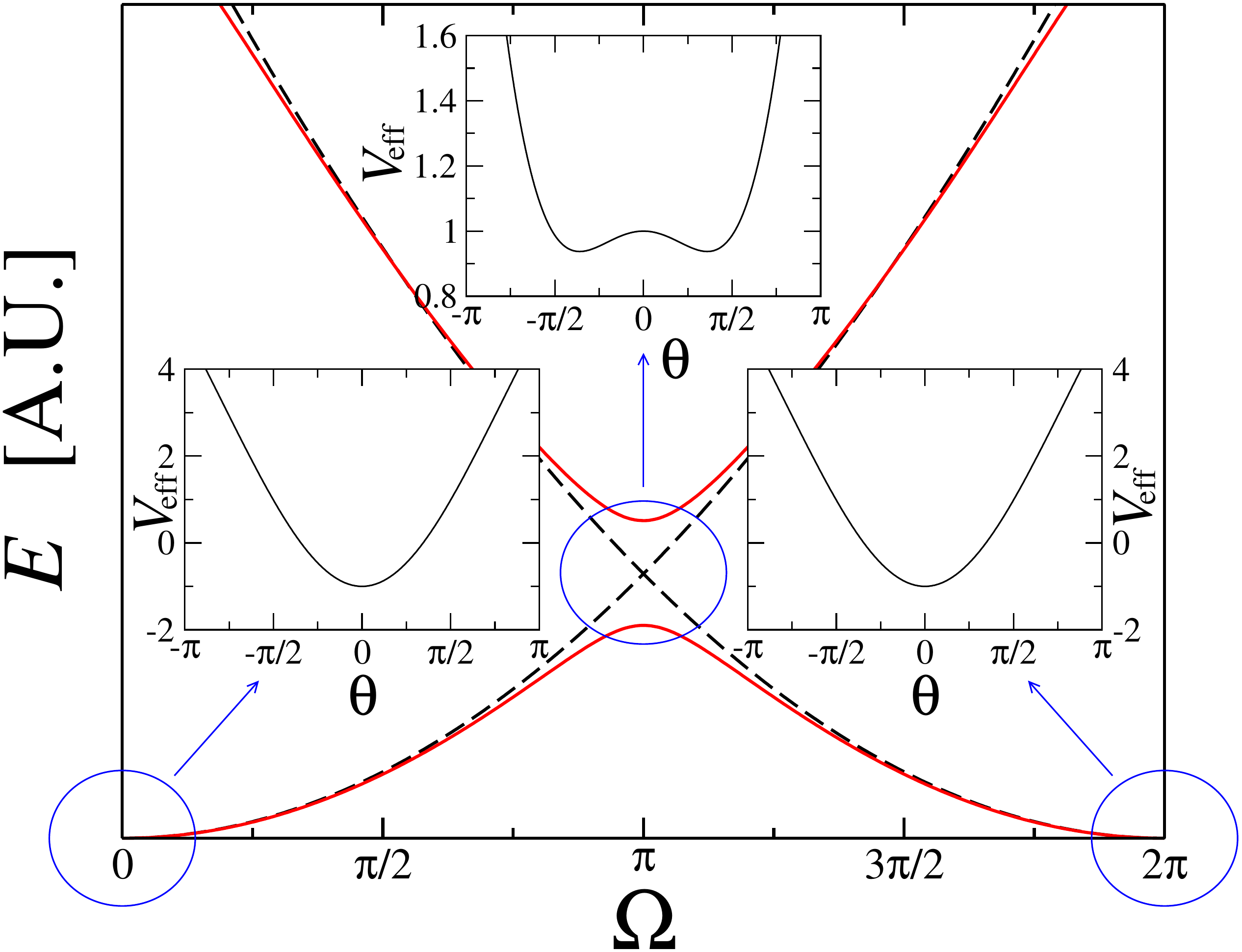}
  \caption{Main panel: sketch of the qubit energy splitting,
    due to the barrier $\Lambda$, for the two lowest-lying energy states
    in the many-body spectrum of model~\eqref{Model}.
    Black dashed lines denote the ground-state energy in the absence of the barrier, as a function
    of the flux $\Omega$. Switching on the barrier opens a gap at the frustration
    point $\Omega = \pi$ (continuous red lines).
    The three insets show the qualitative form of the effective potential
    at $\Omega = 0, \, \pi, \, 2 \pi$.
    Note the characteristic double-well shape forming at $ \Omega=\pi$.
    The qubit, or effective two-level system, corresponds to the two lowest energy levels
    of this potential. In this figure the energies are plotted in arbitrary units.}
  \label{fig:Model_sketch}
\end{figure}

\subsection{Identification of the qubit: effective two-level system}

The Hamiltonian~\eqref{Model} is manifestly periodic in $\Omega$ with period $2\pi$.
Therefore, we can restrict our study to the first rotational Brillouin zone,
(actually to half of it, i.e., $\Omega \in [0,\pi]$,
due to the further symmetry $\Omega \leftrightarrow -\Omega$).
In the absence of a barrier, the system is also rotationally invariant and
therefore the particle-particle interaction energy does not depend on $\Omega$.
The many-body ground-state energy, as a function of $\Omega$, is therefore given
by a set of parabolas each corresponding to a well defined angular momentum state, 
shifted with respect to each other by a Galilean transformation 
and intersecting at the frustration points $\Omega_{j} = (2j + 1)\pi$~\cite{leggett,loss}.
The presence of a finite barrier, $\Lambda>0$, breaks the axial rotational symmetry
and couples different angular momenta states, thus lifting the degeneracy 
at $\Omega_{j}$ by an amount $\Delta E_1$, see Fig.~\ref{fig:Model_sketch}.
The larger $\Lambda$, the larger is $\Delta E_1$, corresponding to the width 
of the gap separating the first two bands.
Provided other excitations are energetically far enough from the two competing ground-states,
this will identify the two-level system defining the desired qubit and its working point.

Below, we discuss this issue with two different approaches:
first, exploiting the mapping of the BH model to the quantum phase model, 
neglecting the fluctuations of the amplitude of the superfluid order parameter; 
this approach can capture, in particular, the regime of a large filling per lattice
site~\cite{Fazio_Rev, AmicoPenna}. 
Then, via numerical calculation of the ground and first three excited energy levels 
of the BH model Eq.~\eqref{Model}, we cover the case of lattice rings with a low filling.

\paragraph{Quantum phase model.}

In the regime of filling much larger than one, the number fluctuations on each site 
can be neglected and the behavior of the system is governed by the 
quantum phase model~\cite{Fazio_Rev, AmicoPenna} with Josephson couplings
$J_j \sim \langle n \rangle |t_j|$, where $\langle n \rangle$ is the average number of bosons per well. 
The presence of a barrier constriction
can be modeled by a weak link $J_{j_0} = J^\prime < J$~\footnote{This model is physically equivalent to the one 
  considered in Eq.(1) being the the barrier modeled 
  as a site-dependent tunnel energy.}
The artificial flux $\Omega$ can be gauged away everywhere but at the $j_0$th
site~\cite{schulz-shastry,Osterloh,eckern},
thus giving rise to an energy term $-J^\prime \cos(\phi_{j_0} - \phi_{j_0+1} - \Omega)$. 
In this situation an effective action can be derived, which depends on a single 
phase difference $\theta=\phi_{j_0} - \phi_{j_0+1}$ across the weak link~\cite{Rastelli,Amico_qubit}.
The corresponding effective potential reads $V_{\rm eff}(\theta) = J \theta^2/M - J' \cos(\theta-\Omega)$,
which, for large $M J'/J$ and moderate $M$, defines a two-level system with degeneracy point at $\Omega=\pi$,
as pictorially illustrated in Fig.~\ref{fig:Model_sketch}.
In the co-rotating frame, these two states correspond to the symmetric and antisymmetric
combination of counter-circulating currents,
where the degeneracy is split because of the inter-well tunneling.

\paragraph{Bose-Hubbard model.}

Here we study the low-lying spectrum of the BH model~\eqref{Model} by a numerical analysis, 
performed in the dilute limit (low filling regime).
This is complementary to the quantum phase model, in that we take into account 
the effect of the number fluctuations, and hence of the amplitude 
of the superfluid order parameter, on the lattice sites.
In Fig.~\ref{twolevels} we show the ED results for $M=16$ and $N=4$.
The top-left panel shows how large interactions and moderate barrier strengths cooperate 
to define a doublet of energy levels at $\Omega=\pi$, well separated in energy 
with respect to the higher excited states; weaker interactions and larger barrier strengths, 
in contrast, do not allow for a clear definition of a two-level system (top-right panel).
We observe that for increasing $\Lambda$, as expected, the gap increases 
and the bands become flatter, thus weakening the dependence of the energy on $\Omega$.
The lower two panels display a complete analysis of the behavior of the spectral gap 
and its distance to the next excited level at $\Omega=\pi$ as a function
of interactions and barrier strength, allowing us to identify the parameter regime 
for the existence of an effective two-level system.
We notice in particular that weakly interacting gases cannot give rise to a sensible qubit 
within this approach, since one cannot isolate two levels out of the many-body spectrum 
with the sole tuning of the barrier strength, while this is possible for larger interaction strengths $U$.

\begin{figure}
\centering
  \includegraphics[width=0.6\textwidth]{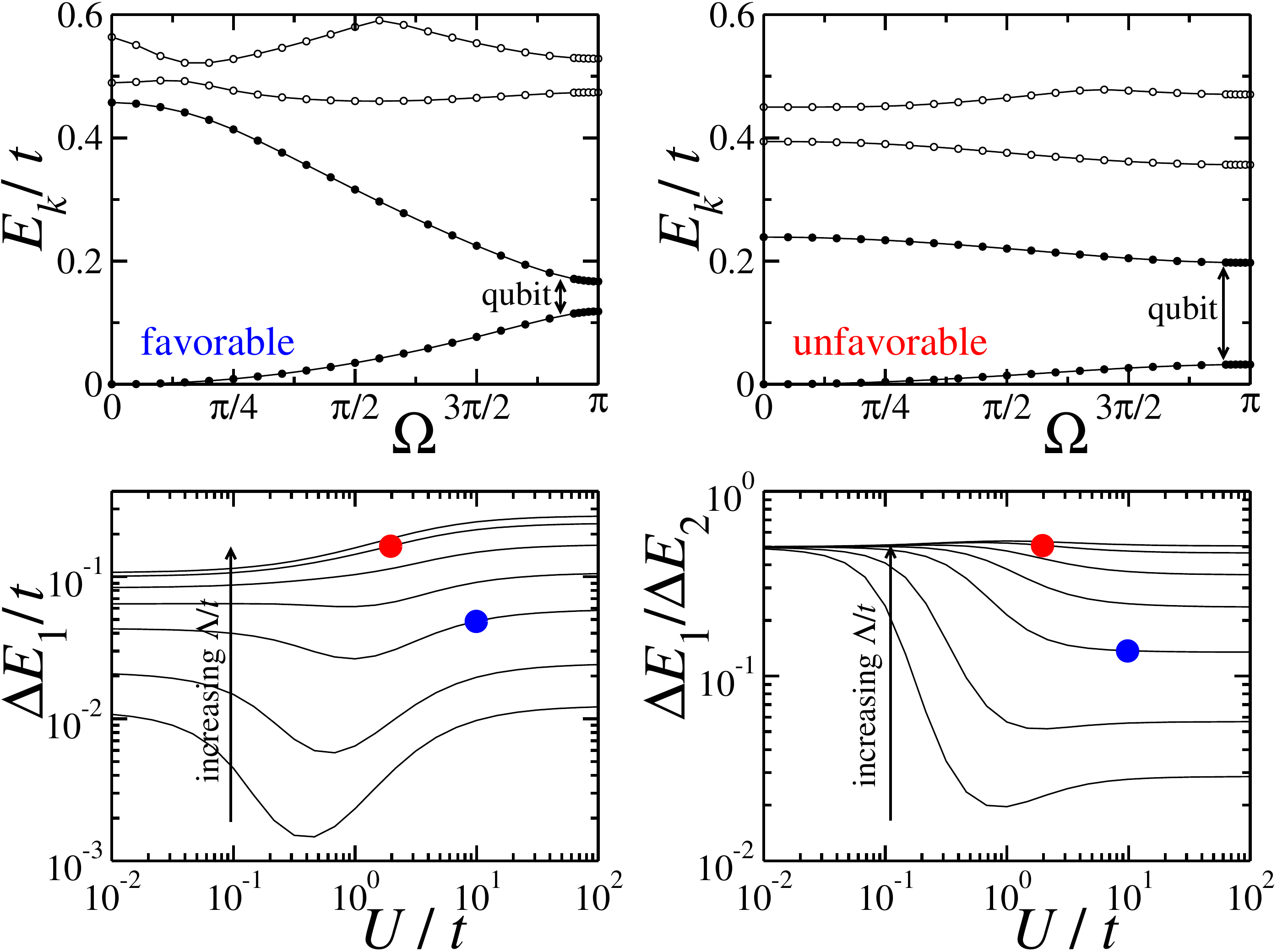}
    \caption{Low-energy spectrum of the BH model for various values of the interaction
    and the barrier strength at fixed size $M=16$ and filling $N/M=1/4$.
    Upper panels: the four lowest energy levels as a function of $\Omega$,
    for $U/t = 10, \, \Lambda /t = 0.5$ (left) and $U/t = 2, \, \Lambda = 5$ (right).
    Lower panels: behavior of $\Delta E_1$ and $\Delta E_1 / \Delta E_2$
    as a function of $U$, for different values of $\Lambda / t$
    (curves from bottom to top: $\Lambda / t = 0.1, 0.2, 0.5, 1, 2, 5, 10$).}
    \label{twolevels}
\end{figure}

Using the above results, we conclude that the low-energy spectrum of the system~\eqref{Model}
may define a qubit over a broad range of lattice filling values.
It is vital for the manipulation of the qubit, though, to explore its quality. 
This implies in particular to study the dependence of $\Delta E_1$ on the system size
and on the interaction strength, as will be considered in the next Section~\ref{Sect:gaps}. 
We will also analyze the nature of the qubit states; this will be the subject of Section~\ref{Sect:Momentum}.

\subsection{Density profiles}

Before presenting our results concerning the quality of the qubit,
we first focus on the density profiles of the gas close to the qubit working point, $\Omega=\pi$.

\begin{figure*}[!t]
\centering
  \includegraphics[width=0.45\textwidth]{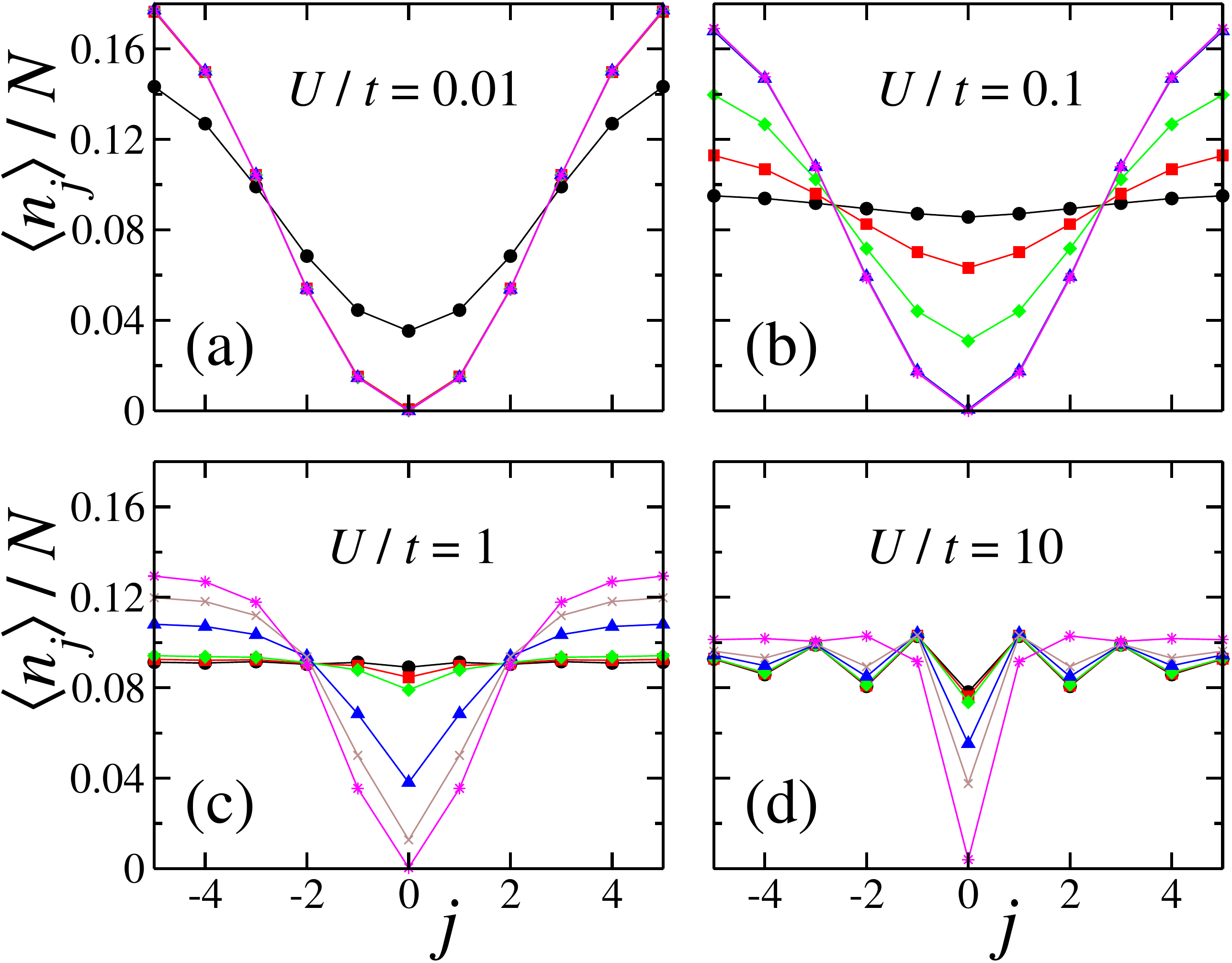}
  \hspace*{1.cm}
  \includegraphics[width=0.44\textwidth]{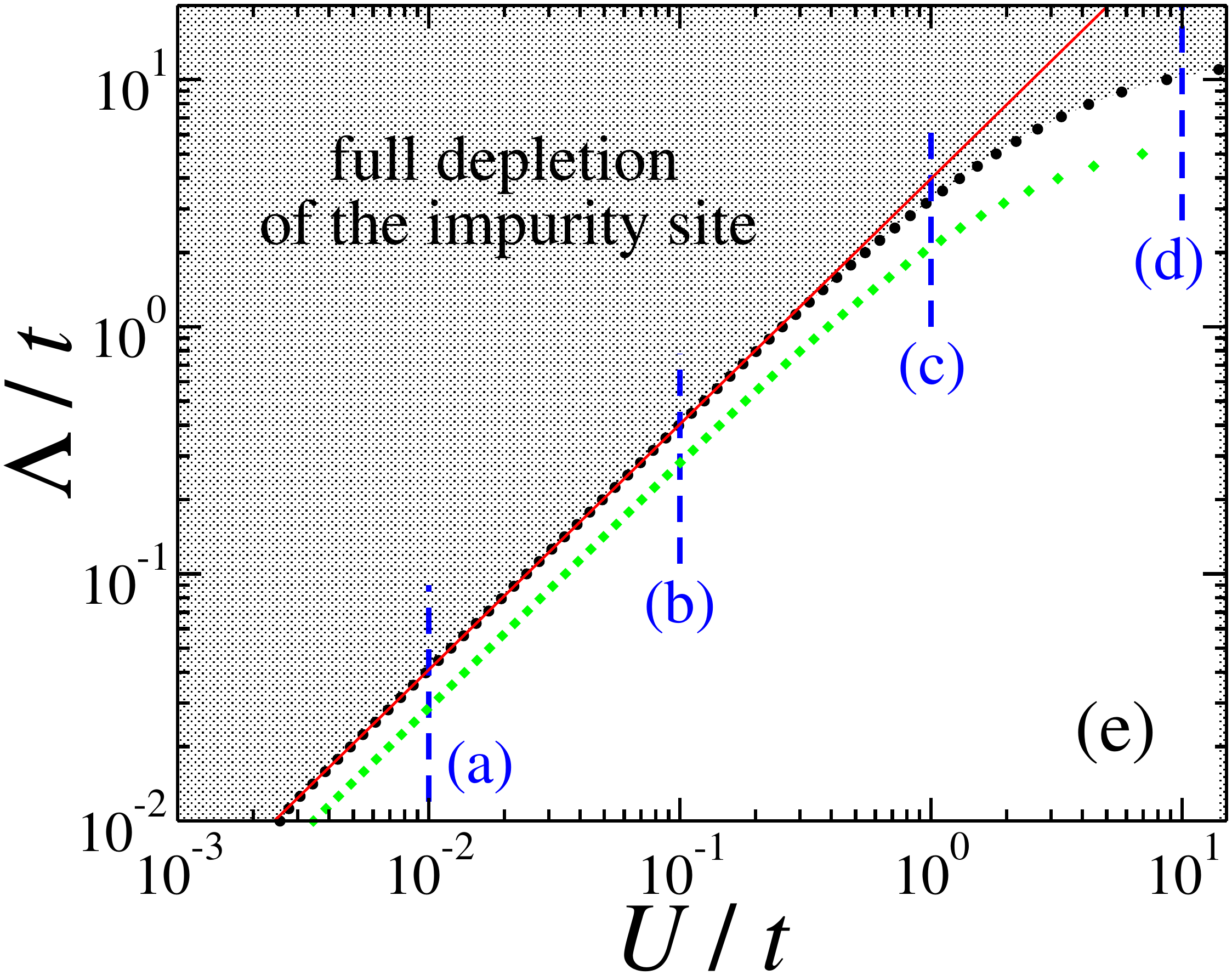}
  \caption{Panels a-d): spatially-resolved density profiles $\langle n_j \rangle /N$
    at $\Omega = \pi$, along a ring with $M=11$ sites and $N=5$ particles,
    for different interaction regimes.
    The various data sets correspond to different values of the barrier strength:
    $\Lambda/t= 0.01$ (black circles), $0.05$ (red squares),
    $0.1$ (green diamonds), $0.5$ (blue triangles), $1$ (brown crosses), $5$ (orange stars).
    Panel e): critical value $\Lambda_c$ as a function of $U$ discriminating
    the parameters region in which the boson density per particle at the
    barrier position is less than the threshold value $\varepsilon = 10^{-3}$
    (black circles refer to $N/M=5/11$, green diamonds are for $N/M=4/16$).
    Vertical dashed lines denote the cuts analyzed in the different left panels,
    for the data set $N/M = 5/11$, while the straight red line is a power-law fit
    $(\Lambda/t)_c \propto (U/t)^{0.99374}$.} 
  \label{dens}
\end{figure*}

An evident effect of the barrier is a suppression of the particle density 
in its immediate proximity; depending on the ring size,
the whole density profile along the ring may well be affected.
The interplay between the interaction strength $U$ and the barrier intensity $\Lambda$
implies different behaviors~\cite{marco}, as exemplified in Fig.~\ref{dens} (panels a-d) 
for relatively small rings. The depth of the density depression increases monotonously 
with $\Lambda$ (inside each panel), while its width decreases with increasing $U$ (see the different panels) 
since the density can be suppressed at the impurity site at the expense of multi occupancy of the other sites;
the latter effect implies a non trivial dependence of the healing length on interaction strength. 
At strong repulsive interactions we also observe small Friedel-like oscillations of the density, 
which are a consequence of the peculiar strong correlations of 1D bosons 
that make their response to impurities similar to fermions.

We note that, a sufficiently large barrier (at fixed $U$) makes the density profile vanish, 
thus effectively disconnecting the ring (panels a-d of Fig.~\ref{dens}). 
The barrier strength required to disconnect the ring depends on the interaction strength. 
Panel e) of Fig.~\ref{dens} shows the result of a thorough analysis of the transition line 
in the $\Lambda$-$U$ plane: for a wide range of interaction strengths, the critical 
barrier height $\Lambda_c$ displays a nearly perfect linear behavior with $U$. 
The prefactor turns out to be nearly proportional to the filling.

\section{Energy gap of the two-level current-flow system}
\label{Sect:gaps}

In this section we study in detail the spectroscopy of the qubit.
We will analyze how the energy gaps $\Delta E_1, \, \Delta E_2$ between the ground and, 
respectively, the first-excited / second-excited energy levels of the many-body 
Hamiltonian~\eqref{Model} depend on the system size and on the filling,
for different $\Lambda$ and $U$.
We find that the qubit is well resolved in the mesoscopic regime of intermediate ring sizes, 
and that it is at best separated from the higher energy levels of the many-body spectrum 
in the regime of strong interactions and weak barrier.

\subsection{Scaling with the system size}
\label{dmrg_scaling}

\begin{figure}[t]
  \centering
  \includegraphics[width=0.6\textwidth]{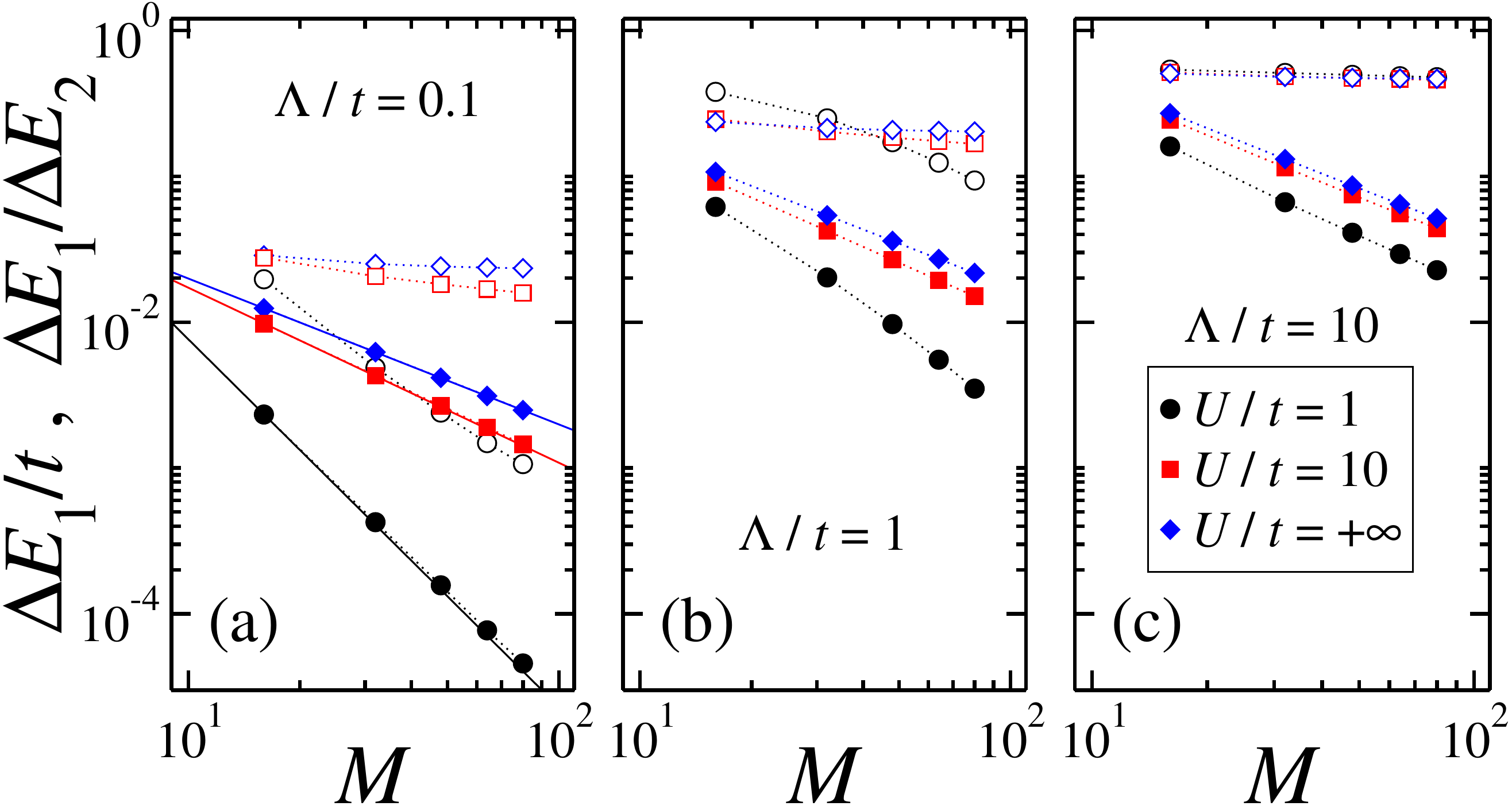}
  \caption{Finite-size scaling of the qubit gap $\Delta E_1$ in units of $t$ (filled symbols)
    and of the ratio between the gaps $\Delta E_1 / \Delta E_2$ (empty symbols),
    at fixed density $N/M = 1/4$.
    Different colors stand for three values of interaction $U/t$, as specified in the legend.
    The various panels are for a fixed barrier $\Lambda/t = 0.1$ (a), $1$ (b), $10$ (c).
    Straight lines in the left panel correspond to the power-law behavior
    predicted by the Luttinger-liquid analysis in the small-barrier limit~\eqref{eq:scaling},
    for the values of the Luttinger parameter $K|_{U=\infty}=1.00$, $K|_{U=10}=1.20$ and $K|_{U=1}=2.52$.}
  \label{scaling}
\end{figure}

In Fig.~\ref{scaling} we show both the qubit gap $\Delta E_1$ and the separation
of the two levels from the rest of the spectrum in terms of $\Delta E_1 / \Delta E_2$,
as obtained by DMRG simulations at constant filling $N/L = 1/4$ (see~\ref{dmrg}).
The three panels correspond to different barrier intensities, from very weak to very high;
each panel containing the three curves at varying interactions from moderate to hard-core.
A clear power-law decay of $\Delta E_1$ results in all the regimes;
the exponents depend on the interplay between the barrier and interaction strengths.

In the small-barrier limit, we can work out the observed scaling law of the gap 
analytically resorting to the Luttinger-liquid effective field theory (see, e.g., Ref.~\cite{marco}).
Indeed we obtain that the quantum fluctuations of the density renormalize the barrier strength
according to $\Lambda_{\rm eff}\sim\Lambda (d/L)^K$,
where $d$ is a short distance cut-off of the low-energy theory, $L= a M$ is the system size, 
$a$ being the lattice spacing, and $K$ is the Luttinger parameter~\cite{marco}.
This yields the scaling of the gap with $M$ as,
\begin{equation}
  \Delta E_1\sim \nu\Lambda_{\rm eff}\sim M^{-K}\;,
  \label{eq:scaling}
\end{equation}
in agreement with the result found in Ref.~\cite{prokofev} for a single impurity potential.
As illustrated in panel a) of Fig.~\ref{scaling}, we find a very good agreement
between the numerical data and the power-law behavior dictated by the Luttinger parameter 
obtained via the Bethe-Ansatz solution of the continuous model (a Lieb-Liniger gas~\cite{LiebLiniger}), 
suitable in the dilute limit of the BH model~\footnote{We have checked as well (not shown) that the same values for $K$, within numerical precision, are extracted from the fit of the decay of the first-order correlation function with the functional form predicted in~\cite{cazalilla} for the finite-size system.}.
For stronger barriers, interestingly, we observe in Fig.~\ref{scaling}(b-c)
that the gap scales again as a power-law, beyond the regime of validity
of the analytical predictions.
We also notice that the scaling of the gap is closely related to the scaling of the
persistent currents flowing along the ring~\cite{Cominotti2014}, which is determined
by the shape of the ground state energy band.

By looking at the separation of the effective two-level system from the rest of the
spectrum (dashed lines in Fig.~\ref{scaling}), we can then start to identify an ideal
regime of size, interaction and barrier for a realistic operational realization of the qubit.
At low barrier intensity $\Lambda/t = 0.1$ (panel a), indeed, a mesoscopic lattice
of few tens of sites filled with mildly interacting bosons
appears to be the best choice, since it would allow for a qubit gap of some $10^{-3} t$,
while this being only a $\simeq 10^{-2}$ fraction of the second excitation energy.
Rings that are too large in size would improve the definition of the two-level system,
yet at the price of too small a resolution of the qubit levels for practical addressing.
When the barrier becomes stronger, the size dependence
of $\Delta E_1 / \Delta E_2$ becomes less and less important, with its absolute value
increasing more and more (i.e., the qubit gets less and less isolated).
Still, at intermediate barrier strengths $\Lambda/t = 1$ (panel b),
a nicely addressable pair of levels with splitting of some $10^{-2} t$, and a relative
separation from the spectrum of order $10^{-1} t$, can be obtained in a mesoscopic
lattice of $M \simeq 16$ sites with relatively weak interactions $U=t$.
Conversely, if the barrier is strong enough to effectively cut the ring,
the low lying levels of the many-body spectrum get almost equally spaced
and therefore the qubit definition results to be poor.

\begin{figure}[!t]
  \centering
  \includegraphics[width=0.6\textwidth]{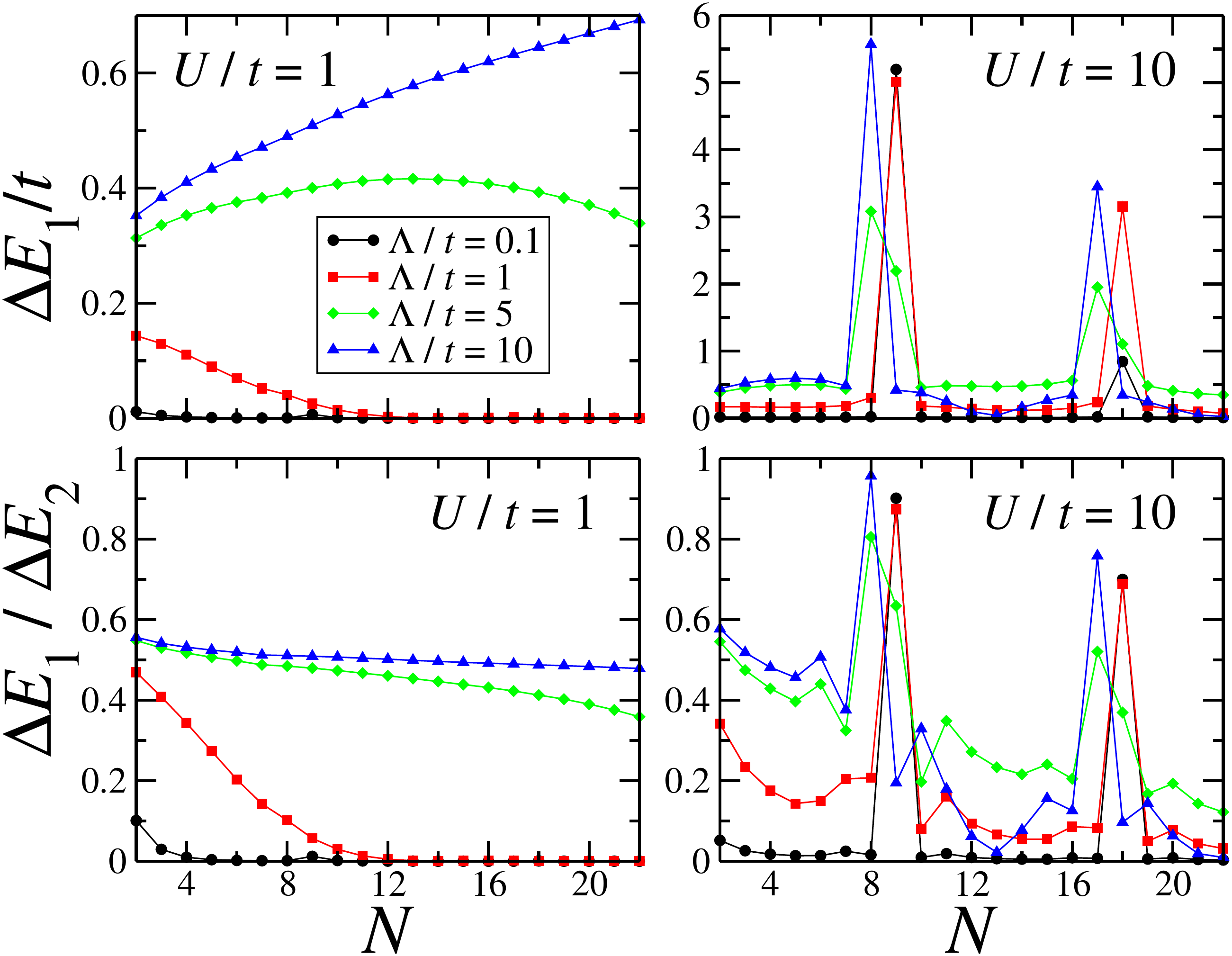}
  \caption{Energy gap $\Delta E_1$ in units of $t$ and the ratio $\Delta E_1/\Delta E_2$
    for $M=9$ lattice sites, at $\Omega=\pi$. 
    We consider the interaction strengths $U/t=1$ (left) and $10$ (right).
    In each plot the various curves stand for $\Lambda/t=0.1$ (black circles),
    $1$ (red squares), $5$ (green diamonds) and $10$ (blue triangles).
    We use an ED technique where, for $N/M > 1$, we allowed a truncation
    in the maximum occupation per site equal to $6$ particles.}
  \label{Gap3}
\end{figure}

\subsection{Dependence of the qubit energy spectrum on the filling factor in mesoscopic rings}
\label{Sect:gap_mesoscopic}

We concentrate next on the mesoscopic regime of few lattice sites, where,
according to our scaling analysis at fixed small filling, the qubit enjoys simultaneously
a clear definition with respect to the other excited states and a good energy resolution.

In Figs.~\ref{Gap3} and \ref{Gap4} we present our results for the gap $\Delta E_1$ and $\Delta E_2$ 
as a function of the filling at fixed system size,
studying its dependence on the barrier and on the interaction strength.
The top panels of Fig.~\ref{Gap3} present the data for fixed interaction strength
($U/t = 1$ and $U/t=10$, respectively)
with the curves representing barrier strengths from weak to strong.
At small $U$ (top-left), we observe a smooth dependence of $\Delta E_1$ on the boson filling, as expected
in the superfluid regime of the Hamiltonian~\eqref{Model}, of which the small ring is reminiscent.
The increase of the barrier strength has two effects: first, at fixed filling, it increases the gap
since it enhances the effect of the breaking of the rotational invariance and therefore lifts the degeneracy at half-flux.
In addition, it changes the dependence of the gap on the filling from being
monotonically decreasing to monotonically increasing,
passing through a crossover situation.
Since the healing length scales as $\xi\propto 1/\sqrt{\nu U}$, at small barrier strengths, a weakly interacting Bose gas screens the barrier, effectively reducing its  height as the density is increased. On the other hand, for a large barrier, the system is effectively in the tunnel limit, and the situation is reversed. The barrier strength is effectively enhanced, since the tunnel energy required to move one particle from one side of the barrier to the other increases if the number of particles or the interaction strength are increased (in order to accommodate the tunnelling particle, the other particles have to readjust their configuration).

At large $U$ (top-right) in Fig.~\ref{Gap3}, $\Delta E_1$ displays a more complex dependence on the filling,
with pronounced peaks at particle numbers commensurate (or quasi) with the size,
related to the presence of Mott lobes in the phase diagram of Hamiltonian~\eqref{Model}~\cite{moreno}.
For weak barrier, indeed, the peaks appear at integer values of $N/M$,
while for very strong potential constrictions the density is suppressed on one site:
the system is close to a lattice with $M-1$ sites and peaks are consequently shifted.
At intermediate barrier strengths we can observe a transient
between the two regimes and broader peaks appear.
Considering the very small system size, this effect arises because the presence 
of the healing length affects the whole bosonic density profile of the ring.

\begin{figure}[!t]
  \centering
  \includegraphics[width=0.6\textwidth]{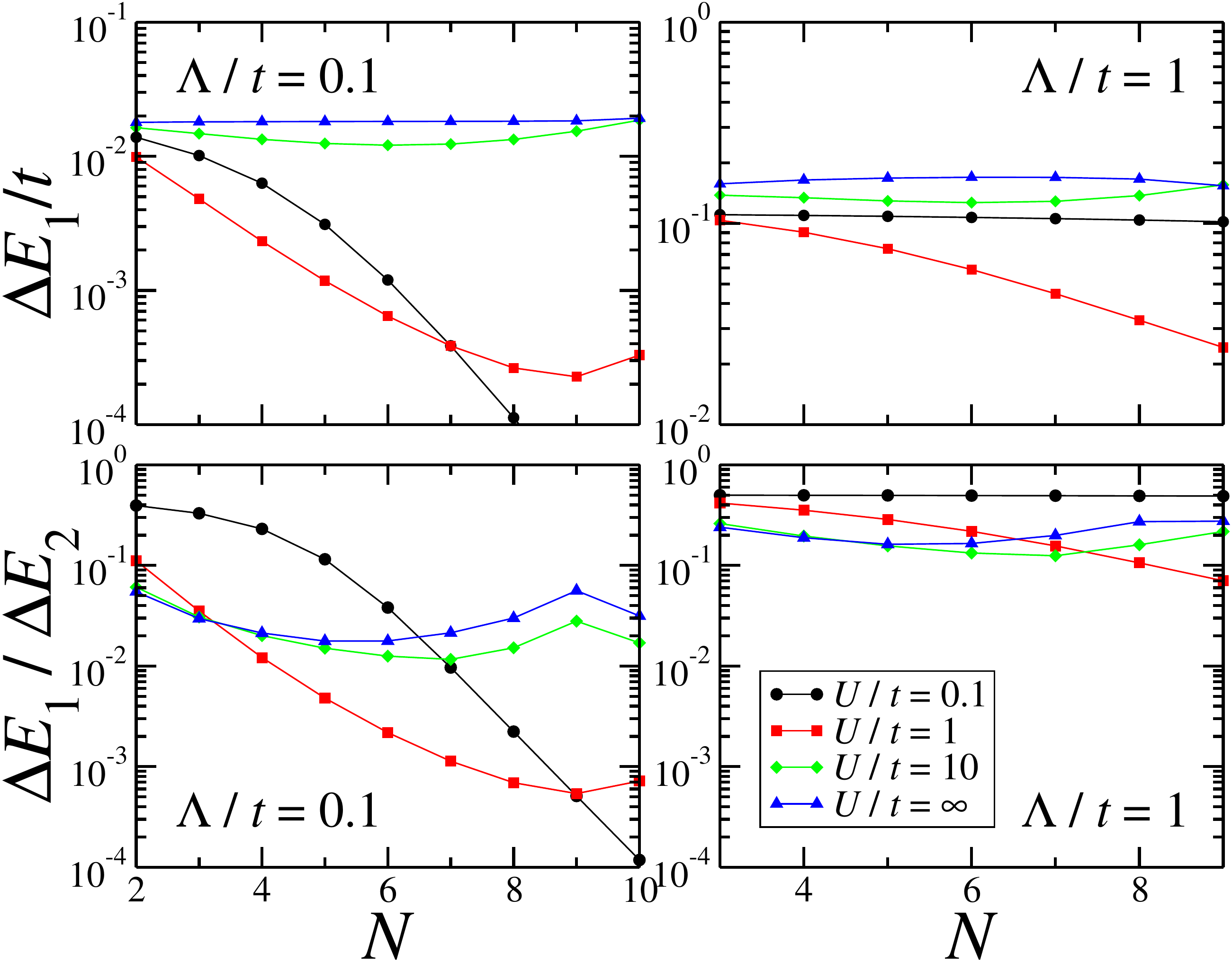}
  \caption{Same as in Fig.~\ref{Gap3}, but for $M=11$ lattice sites
    and for fixed barrier strength $\Lambda/t=0.1$ (left), $1$ (right). 
    The different curves are for $U/t=0.1$ (black circles), $1$ (red squares),
    $10$ (green diamonds) and $\infty$ (blue triangles).}
  \label{Gap4}
\end{figure}

The top panels of Fig.~\ref{Gap4} present data for fixed barrier strength
($\Lambda = 0.1$ and $\Lambda=1$, respectively)
with the curves representing interaction strengths from extremely weak to infinite values.
First, we can clearly see the non-monotonous dependence of the gap on $U$,
which was illustrated in Fig.~\ref{twolevels}, to hold at all fillings in both panels.
Secondly, we notice that the dependence of $\Delta E_1$ on $N$ drastically changes
increasing the interaction strength, displaying different regimes: quickly decreasing,
non monotonous and almost constant. The rapid decrease of the energy gap at weak interactions
can be understood (through a perturbative argument) in terms of level mixing of single-particle energies,
which increases with the number of bosons involved~\cite{dunningham}.
In the opposite regime of hard core bosons, the energy gap is of the same order as the one
of the non-interacting Fermi gas. This can be readily understood in terms of the TG Bose-Fermi mapping:
indeed, in a non-interacting Fermi gas the energy gap is given by
$\Delta E_{1}=(\sum_{j=1}^{N-1}\epsilon_{j}+\epsilon_{N+1})-\sum_{j=1}^{N}\epsilon_{j}$,
where $\epsilon_j$ are the single-particle energies.
In particular, for a small barrier, using perturbation theory, one obtains that
the single-particle energy gaps $\epsilon_{j+1}-\epsilon_{j}$ are identical
for all the avoided levels crossings, hence the gap $\Delta E_1$ is independent of the filling.

The lower panels of Figs.~\ref{Gap3} and~\ref{Gap4} display the ratio $\Delta E_1/\Delta E_2$. 
This allows us to identify the low-barrier, intermediate-to-large interaction regime 
at arbitrary filling as the most favorable for the qubit. 
Indeed, depending on the interaction strength, a too large barrier yields an unfavorable situation 
similar to the one depicted in the top-right panel of Fig.~\ref{twolevels}, 
where $\Delta E_2 \sim \Delta E_1$. It is interesting to notice that these unfavorable cases 
correspond to values of barrier and interaction strength in the right panel of Fig.~\ref{dens} 
where the ring is effectively disconnected. 
This allows us to identify the ratio $\Lambda/U$ as a useful parameter to define the quality 
of the qubit in terms of its energy resolution: the most advantageous parameter regime 
for the qubit corresponds to the lower half-plane in Fig.\ref{dens} (e), below the critical line.

In summary, this analysis shows that a particularly favorable regime for the energy resolution 
of the qubit is the Tonks-Girardeau and small-barrier limit, where the system has a well defined gap,
independent of the particle number and well separated from the remaining part
of the many-body spectrum. However, for the realization of a tunable-gap qubit,
the limits of weak interaction with low filling
and intermediate interaction with high filling can be useful.

We close this section providing  the order of  magnitude for the gaps discussed above.  For a $^{87}$Rb gasin a mesoscopic ring shaped deep optical lattice of $\sim 50\mu$m circumference and $10$ lattice wells, the hopping energy is of the order of $t\sim 0.5$kHz. This yields a typical energy scale for the gap of tens to few hundreds of Hz, depending on the choice of barrier strength, well within the range of experimental accessibility.

\section{Momentum distributions}
\label{Sect:Momentum}

So far we focused our analysis on the behavior of the energy spectrum of the qubit 
as a function of the system parameters. We now investigate the ground state of the system in more detail. 
Special care is devoted to the regimes corresponding to a macroscopic superposition of circulation states.
We assess the {detectability} of the latter through the study of the momentum distribution.

The momentum distribution is experimentally accessible in cold atoms experiments
via TOF expansion measurements, {by averaging over many repeated TOF realizations~\cite{TOF_review,pitaevskii}}, and is employed to get information about the current circulation along the ring~\cite{solenov,moulder,murray}. It is defined as
the Fourier transform with respect to the relative coordinate of the one-body density matrix
$\rho_{(1)}(\textbf{x},\textbf{x}')=\langle\hat{\psi}^{\dagger}(\textbf{x})\hat{\psi}(\textbf{x}') \rangle$:
\begin{equation}
  n(\textbf{k})=\int \mbox{d}\textbf{x}\int \mbox{d}\textbf{x}' \langle\hat{\psi}^{\dagger}(\textbf{x})\hat{\psi}(\textbf{x}') \rangle e^{i\textbf{k}\cdot(\textbf{x}-\textbf{x}')},
  \label{nk}
\end{equation}
where $\textbf{x}$ and $\textbf{x}'$ denote the position of two points
along the ring's circumference. 
Although, in general, $\textbf{k}$ is a three dimensional wave vector, here we restrict 
to consider a TOF picture along the symmetry axis of the ring, 
and therefore two dimensional $\textbf{k}$'s.
To adapt Eq.~\eqref{nk} to our lattice system, we use
$\hat{\psi}(\textbf{x})=\sum_{j=1}^{M}w_{j}(\textbf{x})\hat{b}_{j}\;,$
where $w_{j}(\textbf{x})=w(\textbf{x}-\textbf{x}_{j})$ is the Wannier function localized
on the $j$-th lattice site, and $\textbf{x}_{j}$ denotes the position of the $j$-th lattice site.
Thereby, Eq.~\eqref{nk} can be recast into
\begin{equation}
  n(\textbf{k})=|\tilde{w}(\textbf{k})|^{2}\sum_{l,j=1}^{M} e^{i\textbf{k}\cdot(\textbf{x}_{l}-\textbf{x}_{j})} \langle \hat{b}^{\dagger}_{l} \hat{b}_{j} \rangle,
  \label{nkw}
\end{equation}
where $\tilde{w}(\textbf{k})$ is the Fourier transform of the Wannier function.

To avoid effects of the proximity of the superfluid-insulator transition,
in the following analysis, we focus on incommensurate fillings 
(see Section~\ref{Sect:gap_mesoscopic} for a more detailed discussion).

\begin{figure*}[!t]
  \centering
  \includegraphics[width=0.85\textwidth]{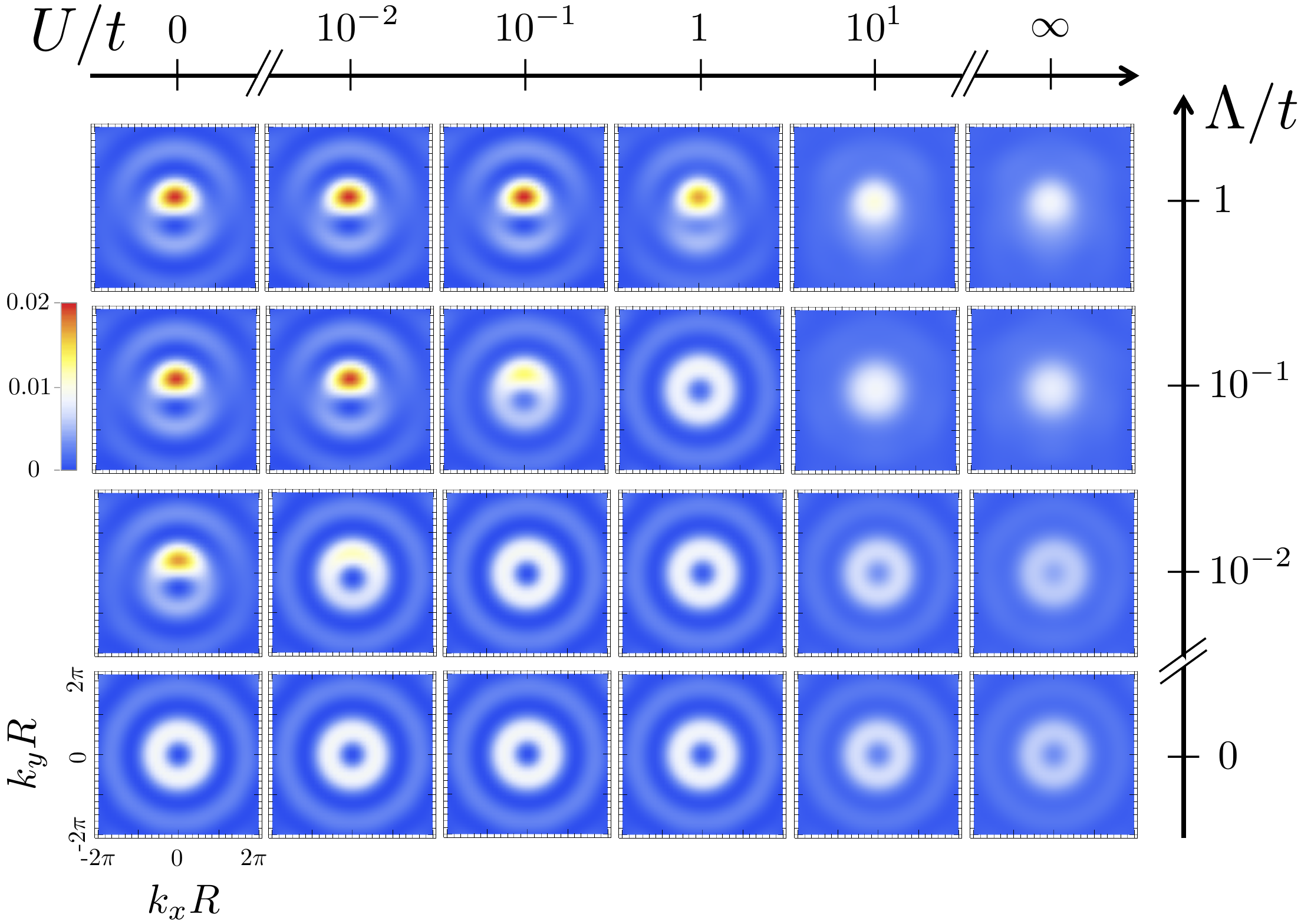}
  \caption{Ground state momentum distribution (TOF) close
    to the degeneracy point: $\Omega=\Omega^+=\pi+\epsilon$
    (hereafter we fix $\epsilon = 10^{-3}$). 
    For $\Omega=\Omega^+$, the ground state corresponds
    to a symmetric superposition of the flow states with zero and one quanta of circulation.
    The superposition depends on the interplay between $U$ and $\Lambda$:
    lines $\Lambda/t=0, \, 0.01, \, 0.1, \, 1$; columns $U/t=0, \, 0.01, \, 0.1, \, 1, \, 10, \, \infty$.
    For the filling value, $N/M=5/11$, used in these graphs, larger values of $\Lambda$
    yield TOF images very close to those for $\Lambda/t=1$
    The results were obtained with the exact diagonalization.}
  \label{tofwithl}
\end{figure*}
In  absence of  barrier $\Lambda = 0$, the system  has no circulation for $\Omega<\pi$ and one quantum of circulation for $\Omega>\pi$, while at the frustration point $\Omega=\pi$, it is a perfectly balanced superposition of the two states. As a consequence, the momentum distribution is peaked at $\textbf{k}=0$ for $\Omega<\pi$ and is ring-shaped for $\Omega>\pi$, as discussed in~\ref{tof-nobarrier}. At $\Omega=\pi$, instead, it displays an interference of the two situations, reflecting the coherent superposition of the two states (see ~\ref{tof-nonint-l}).
When $\Lambda \neq 0$, the superposition state occurs for a wide range of  $\Omega$, thereby displaying interference effects as shown in Fig.~\ref{tofwithl}. 
The relative weight of the two-quanta-of-circulation components in the superposition strongly depends on $\Omega$, $\Lambda$, and $U$. In particular, at the frustration point, the superposition is perfectly balanced, independently of $\Lambda$, and $U$. Away from the frustration points, the relative weights tend to the unperturbed ones carrying zero or one quantum of angular momentum. This phenomenon occurs  over a distance in $\Omega-\pi$ that depends on $\Lambda$: the smaller is $\Lambda$, the faster the unperturbed weights are recovered. For this reason, in Figs.\ref{tofwithl} and \ref{tof_nu},
we slightly off-set  $\Omega$  from the frustration point (the weights of the circulating states are not equal, yet close enough to ensure that both angular momentum states contribute significantly to the superposition). For $\Omega>\pi$, the component carrying one quantum of angular momentum has a larger weight in the superposition, making the effect of the barrier and its screening easily {detectable} in the  TOF image; the opposite situation occurs for $\Omega < \pi$. The TOF results shown in Fig.~\ref{tofwithl} and~\ref{tof_nu} quantitatively depend on the choice of $\Omega$, but the  screening effect of the barrier and  the {detectability of the superposition} result weakly affected.

To understand the TOF results of Fig.~\ref{tofwithl}, it is instructive to consider first the case
without interactions, $U=0$, that is analytically accessible.
The corresponding momentum distribution close the frustration point and for a weak barrier
reads (see Eq.~\eqref{eq:nk} in~\ref{tof-nonint-l} for the derivation)
\begin{eqnarray}
  n(\textbf{k})&=&\sin^{2}(\varphi/2) J^{2}_{n}(|\textbf{k}|R)+ \cos^2(\varphi/2) J^{2}_{n+1}(|\textbf{k}|R) \nonumber \\ 
  &&+ \sin(\varphi) \cos(\gamma_\textbf{k})J_{n}(|\textbf{k}|R)J_{n+1}(|\textbf{k}|R) \;,
  \label{sup_U0}
\end{eqnarray}
where the Bessel functions $J_{n}$ correspond to states with angular momentum $n$,
and $\gamma_\textbf{k}$ is the angle along the ring;
the parameter $\varphi$ is a function of the flux and the barrier strength
(see Eq.~\eqref{eq:phi} in~\ref{tof-nonint-l}).
Eq.~\eqref{sup_U0} shows that the TOF images allow to visualize the superposition between of states
with different angular momenta: the functions $J_0$ and $J_1$ interfere, giving rise
to {\it a peak at zero ${\bf k}$ and a fringe with ring-shaped symmetry}.
The {detectability} of this feature increases with the barrier strength $\Lambda$.
Note that the angular position of the peak in momentum space depends on the position
of the barrier in real space along the ring;
it would be affected by a phase shift between the two states of well-defined angular momentum.

The superposition state for small $U$ can be analyzed in a similar way.
We note in Fig.~\ref{tofwithl} that, for sufficiently weak interactions,
an angular modulation of the ring-shaped momentum distribution arises.
A stronger barrier makes the angular asymmetry increasing,
while the interaction strength, by screening the barrier, leads to the opposite phenomenon.

Upon increasing the interaction strength from intermediate to very large,
we observe a smearing of the modulated ring shape TOF images.
This is an effect of increased quantum fluctuations, which leads,
for strong barrier strengths, to a single maximum centred at non-zero $\textbf{k}$ values.

The very different TOF images between the regimes of weak and strong interactions
can be understood by recalling the different nature of the superposition state
in the various interaction regimes~\cite{hallwood_homogeneous,rey_non-homo}. For instance, at zero or very weak interactions, within the GP regime, the many-body state is a coherent state of single particle superpositions. Increasing the interaction strength to the intermediate regime the superposition
is described by the so-called NOON state $|N,0\rangle+|0,N\rangle$,
i.e., a macroscopic superposition of states where all bosons occupy
either the state with zero angular momentum or the one carrying one quantum of angular momentum. For increasing interactions this many-body entangled state matches the known macroscopic superposition of Fermi spheres at very large interactions~\cite{rey_non-homo,schenke}.

For all regimes of interactions, we notice that the TOF images become independent 
of the barrier above a critical value of the barrier strength, which well agrees
with the critical value $\Lambda_c$ for disconnecting the ring, as identified in Fig.~\ref{dens}. 
Globally, we observe that good-quality TOF images allowing to easily identify 
the superposition of current states as a modulated ring structure are found 
for a ratio $\Lambda/U$ in the vicinity or above the critical line of Fig.~\ref{dens}(e).

\begin{figure}[t]
  \centering
  \includegraphics[width=0.45\textwidth]{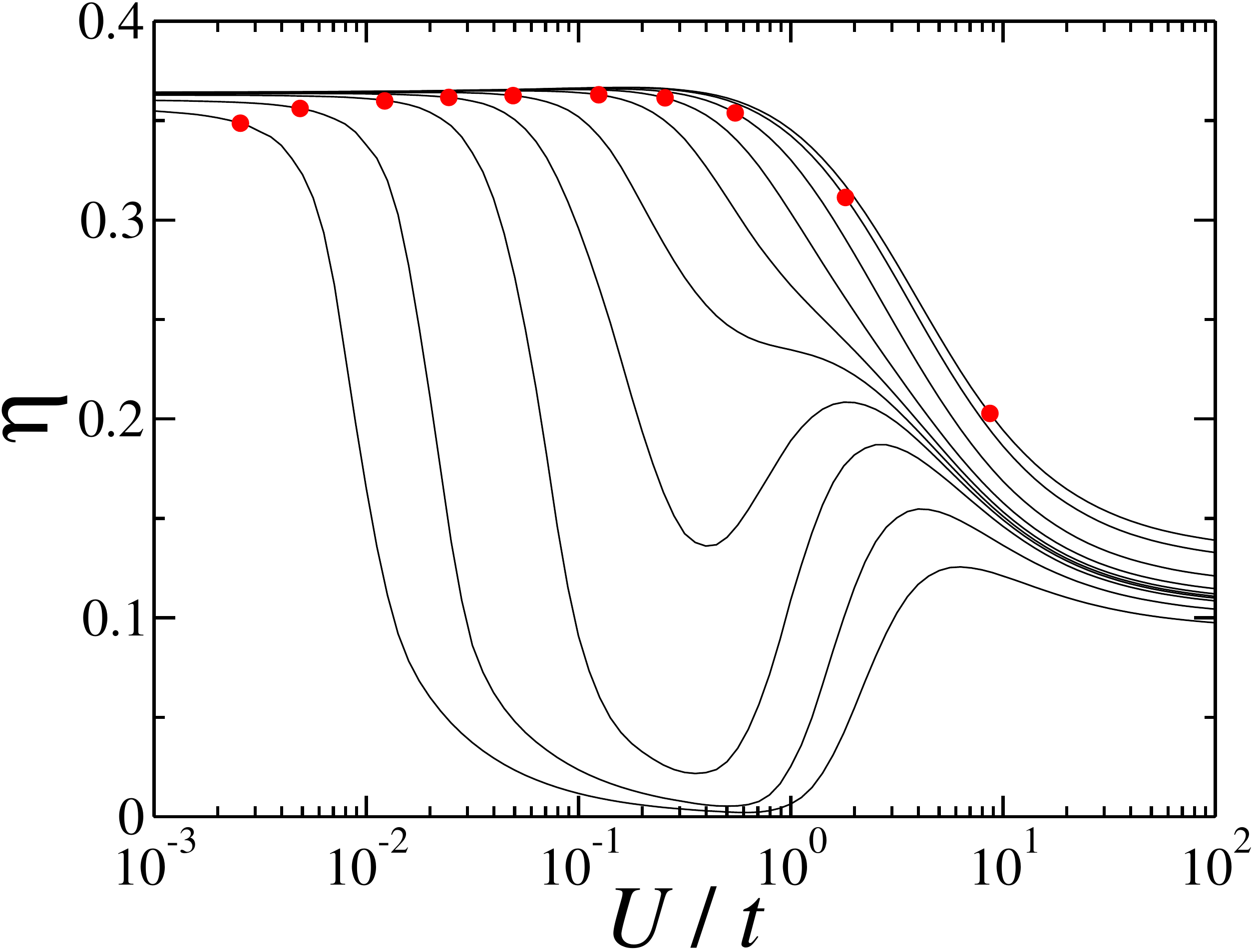}
  \caption{{Averaged contrast} $\eta$ vs interaction strength $U/t$ for different values of the barrier strength
    (curves from left to right: $\Lambda/t=0.01, 0.02, 0.05, 0.1, 0.2, 0.5, 1, 2, 5, 10$).
    The red circles denote the value of $\eta(U_c/t)$, for each value of $\Lambda/t$,
    where $U_c/t$ has been defined from the analysis of Fig.~\ref{dens}.}
  \label{fig:visibility}
\end{figure}

To quantify the {detectability} of the superposition state in the momentum distribution 
for different barrier and interaction strengths, we define the {averaged contrast between the momentum distribution with and without barrier}
\begin{equation}
  \eta=\frac{\int\mbox{d}\textbf{k}\;|n_{\Lambda\neq0}(\textbf{k})-n_{\Lambda=0}(\textbf{k})|}{\int\mbox{d}\textbf{k}\;n_{\Lambda\neq0}(\textbf{k})+n_{\Lambda=0}(\textbf{k})}\;,
\end{equation}
reflecting the modification in the integrated momentum distribution due
to superposition of states induced by the barrier.
We find that $\eta$ is non-monotonic upon increasing the interactions between
the particles, while keeping fixed the barrier strength---Fig.~\ref{fig:visibility}.
This is an effect  of the non-monotonic screening of the barrier as a function of interaction strength,
first predicted in Ref.~\cite{marco} through the study of the persistent-current amplitude.

\begin{figure}[t]
\centering
  \includegraphics[width=0.5\textwidth]{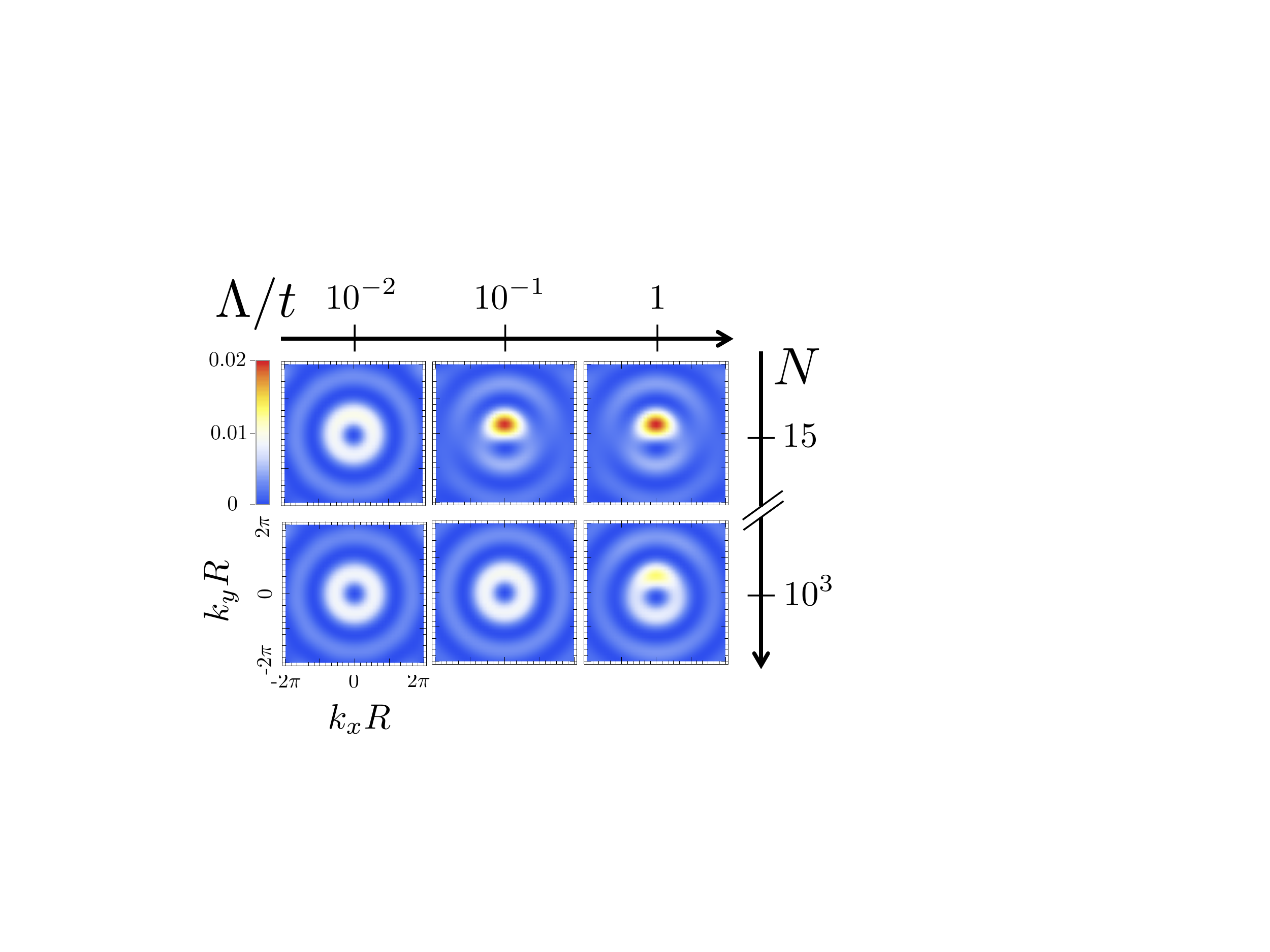}
   \caption{Ground state momentum distribution (TOF) close
     to the degeneracy point: $\Omega=\Omega^+=\pi+\epsilon$ for $U/t=10^{-2}$, 
     $M=11$ and $N=15, 10^3$ (obtained from truncated ED and GP respectively).
     The TOF pictures are qualitatively similar to the ones in Fig.~\ref{tofwithl},
     but the features of the superposition appear at larger values of $\Lambda$,
     compared to the case at lower filling.}
  \label{tof_nu}
\end{figure}

Finally, we comment on the expected behavior for a system with filling larger than one.
In Fig.~\ref{tof_nu} we show the TOF images for larger fillings, ranging from values
of $N/M$ close to one, obtained with truncated ED, to fillings much larger than one,
obtained solving the GP Eq.~\eqref{eq:gp1}.
In both cases we note that the TOF images are qualitatively the same as the ones shown
in Fig.~\ref{tofwithl}, and therefore our analysis is relevant also for systems
with larger number of particles, like the ones employed in the experiments so far.
We notice that, at higher filling, a larger barrier strength is needed, with respect to the lower filling case, to produce the same superposition and to observe the same TOF.

\section{Conclusions}
\label{RingsVsContinuous}

We considered a system of bosonic atoms loaded in a ring-shaped 1D optical lattice potential, 
hosting a localized barrier on a given site of the lattice.
A ring lattice could be produced, for example, through protocols based
on interference patterns of Laguerre-Gauss or Bessel laser beams~\cite{luigi,ring_Franke}
or through Spatial Light Modulators~\cite{Amico_qubit}.
Besides its possible exploitation in quantum technology, the system provides a paradigmatic arena 
to study the interplay between quantum fluctuations, interactions and the role of the barrier potential.

In order to address the qubit effective quantum dynamics encoded into the system, its coherent superposition of current states 
and the scaling of its properties with system size, we considered ring sizes ranging 
from few lattice sites ($10-20$) to larger structures ($100$).
We provide a direct evidence that the qubit dynamics can be achieved
beyond the pure superfluid phase dynamics conditions (described by the quantum phase model)
exploited to derive the effective double-well potential~\cite{Amico_qubit},
see Figs.~\ref{fig:Model_sketch} and~\ref{twolevels}.
Incidentally, we note that one- and two-qubit gates can be realized with our quantum device 
by tuning the the barrier height and interaction suitably (see Ref.~\cite{Amico_qubit} for details).

We studied the resolution and {detectability} of the qubit, by following two routes:
the analysis of the scaling properties (both with the number of particles and system size)
of the energy gap between the two energy states, and the analysis of the momentum distribution.

{\it The energy gap of the two-level system.}
We quantified the scaling of the gaps $\Delta E_1 $ and $\Delta E_2 $ with system size.
Our results indicates that $\Delta E_1$ is appreciable for small and mesoscopic systems
while it is suppressed in thermodynamic limit (Fig.~\ref{scaling}), decaying as a power law
with system size. This follows from the localization of the barrier to a point like
on the scale of the lattice spacing, and shows that the lattice potential along the ring bears
several added values with respect to the uniform ring case or the case
of a broad barrier~\cite{rey_non-homo}, ultimately facilitating
the exploitation of the device in future atomtronic integrated circuits.
Our scaling analysis allows us to identify the mesoscopic regime as the most suited
for the realization of the AQUID---see Fig.~\ref{scaling}.

{\it Momentum distribution.} 
For the mesoscopic structures, we demonstrated that the {coherent superposition}
of forward and backward scattering of the particles through the barrier site,
is indeed {detectable} through time-of-flight expansion.
This is a strong {\it a posteriori} evidence of the two-level-system effective physics
that is encoded into the system.
For fixed values of the filling parameter, the {detectability} of the superposition depends
on the relative size between barrier and interaction strengths:
the barrier makes the {detectability} increasing, while the interaction strength, screening
the barrier, leads to the opposite phenomenon, yielding a non-monotonous behavior of the {detectability and averaged contrast}.
By increasing the filling parameter for fixed $U$ and $\Lambda$, the screening
of the barrier is enhanced, and therefore the barrier is less effective
in creating the coherent superposition of flows.
A separate discussion is needed for the regime of large interactions.
This regime is characterized by fermionic effects due to strong correlations. 
In particular, we note a good {detectability} of the superposition state with a simultaneous presence 
of density modulations along the ring.

In summary, our work indicates that a {\it mesoscopic} ring lattice with localized barrier 
provides a candidate for the AQUID. 
The qubit dynamics is {detectable} in a wide range of system parameters. 
We have identified the ratio $U/\Lambda$ as an important parameter to discuss the behavior 
of the qubit both in terms of its definition with respect to the rest 
of the many-body spectrum (the qubit turns out to be best defined below the critical line 
in Fig.~\ref{dens}(e)) and its {detectability} in TOF images (the {contrast} is found to be best defined 
around or above the same critical line). 
This allows us to conclude that the ratio $U/\Lambda$ on the critical line 
yields an optimal parameter choice, corresponding to the best trade-off between 
the momentum distribution {detectability} and the gap resolution. Given the flexibility achieved in the actual experiments, we believe that such an optimum parameter regime is well within the current experimental know-how of the field.

\section*{Acknowledgments}

D. A. is indebted to S. Vinjanampathy and P. N. Ma for help implementing the exact diagonalization algorithm.
We thank F. Auksztol, H. Crepaz, and R. Dumke for discussions.
We acknowledge support from the Merlion project `LUMATOM',
the Institut Universitaire de France, the ERC Handy-Q grant N.258608, the ANR project Mathostaq ANR-13-JS01-0005-01, and the Italian FIRB project RBFR12NLNA.
DMRG numerical simulations have been performed on the MOGON cluster of JGU-Mainz.

\appendix

\section{Momentum distribution for $\Lambda=0$ and various interaction strengths}
\label{tof-nobarrier}

The signature of a non vanishing current flow along the ring lattice is a ring-shaped configuration
of the momentum distribution (see~\ref{tof-nonint-l} for a derivation in the non-interacting limit).
Fig.~\ref{tofnol} shows the predicted TOF images in the absence of the barrier
for various interaction strengths. The perfect ring shape reflects angular momentum
conservation at all interaction strengths, consistent with Leggett's theorem
establishing that the persistent currents through a rotationally invariant system
are not affected by the interactions. We note, however,
that the {detectability} in the time-of-flight images is reduced at large interactions,
due to the enhanced role of phase fluctuations.

\begin{figure}[hb]
\centering
  \includegraphics[width=0.5\textwidth]{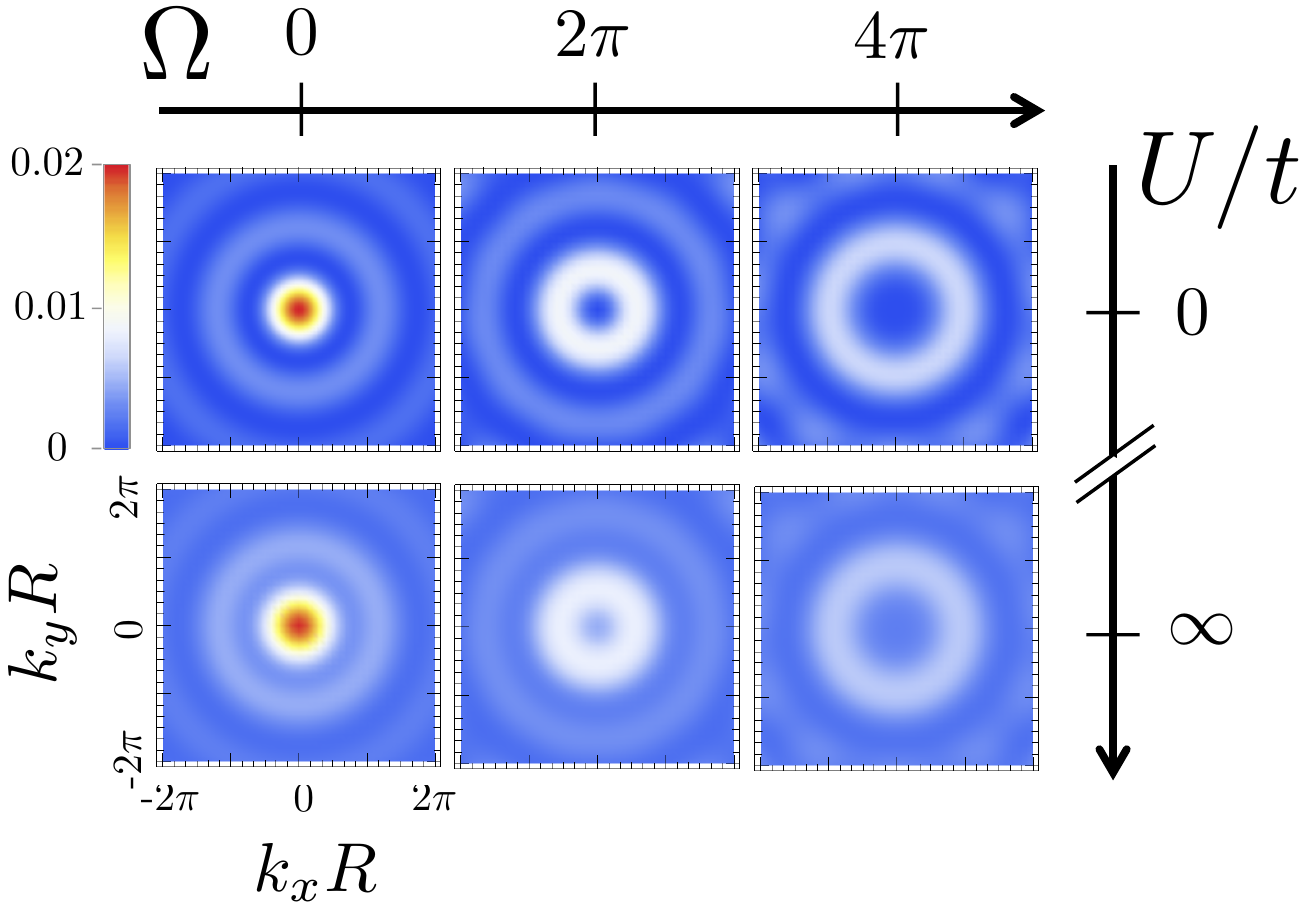}
  \caption{Ground state momentum distribution (TOF) in the absence
    of the barrier, for different values of the Coriolis flux: $\Omega=0, 2\pi, 4\pi$
    and different regimes of interaction strength: non-interacting (upper line) 
    and infinite interactions for $N=5$ (lower line).}
  \label{tofnol}
\end{figure}

\section{Momentum distribution for $U=0$ and arbitrary barrier strength}
\label{tof-nonint-l}

In the noninteracting regime the many-body problem reduces to a single-particle one.
In the absence of the barrier the Schr\"odinger equation, in polar coordinates
and scaling the energies in units of $E_0=\hbar^2/2mL^2$, with $m$ being the particle mass, 
and $L$ the system size, reads
$$\left(-i\frac{\partial}{\partial \theta}-\frac{\Omega}{2\pi}\right)^{2}\psi(\theta)=E\psi(\theta)\;,$$
where $\theta\in[0,2\pi]$.
The wavefunction for a state with defined angular momentum is a plane wave
$\psi(\theta)=(1/\sqrt{2\pi})e^{in\theta}$, where $n\in\mathbb{Z}$ to satisfy
periodic boundary conditions, and the corresponding spectrum is $E_{n}=(n-\Omega/2\pi)^2$.
The momentum distribution then reads
\begin{eqnarray}
  \nonumber n(\textbf{k}) &=& \int\mbox{d}\textbf{x}\int\mbox{d}\textbf{x}'e^{i\textbf{k}\cdot(\textbf{x}-\textbf{x}')}\psi^{*}(\textbf{x})\psi(\textbf{x}')\\
  \nonumber &\sim&\left|\int_{0}^{2\pi}\mbox{d}\theta\;e^{i(k_{x}R\cos\theta+k_{y}R\sin\theta)}\psi^{*}(\theta) \right|^2\\
  &=&\left|e^{im\gamma}J_{m}(|\textbf{k}|R) \right|^2=\left|J_{n}(|\textbf{k}|R) \right|^2\;,
  \label{Bessel}
\end{eqnarray}
where $R=L/2\pi$ is the ring radius, we have defined $\gamma$ as
$k_{x}=|\textbf{k}|\sin\gamma$, $k_{y}=|\textbf{k}|\cos\gamma$, and $J_{n}$ is the $n$-th
order Bessel function of the first kind. For $n=0$ the momentum distribution
is peaked at $\textbf{k}=0$, while for $n>0$ it is ring shaped, with a radius that grows with $n$.

In the presence of a localized barrier of strength $\lambda$ the Schr\"odinger equation becomes
$$\left(-i\frac{\partial}{\partial \theta}-\frac{\Omega}{2\pi}\right)^{2}\psi(\theta)+\lambda\delta(\theta)\psi(\theta)=E\psi(\theta)\;.$$
The effect of the $\delta$-barrier is to mix states with different angular momentum.
For a small barrier we can reduce to the simplest case of mixing of states
that differ by just one quantum of angular momentum, and apply degenerate perturbation theory.
We write the Hamiltonian in the following form
\begin{equation}
  H=\left(
  \begin{array}{cc}
    E_{n} & \lambda/2\pi\\
    \lambda/2\pi & E_{n+1}
  \end{array}
  \right)\;;
\end{equation}
the corresponding eigenvalues and eigenvectors reads:
$$\epsilon_{1,2}=\frac{E_{n+1}+E_{n}}{2}\pm\frac{\sqrt{\delta E^{2}+\lambda^{2}/\pi^{2}}}{2}\;,$$
where $\delta E=E_{n+1}-E_n$, and

\begin{equation}\nonumber
w_{1}=\left(
\begin{array}{c}
\sin(\varphi/2)\\
\cos(\varphi/2)
\end{array}
\right)\;,\;\;\;\;\;\;\;\; w_{2}=\left(
\begin{array}{c}
\cos(\varphi/2)\\
-\sin(\varphi/2)
\end{array}
\right)\;,
\end{equation}
where
\begin{equation}
\cos^{2}(\varphi/2)=\frac{\sqrt{\delta E^{2}+\lambda^{2}/\pi^2}-\delta E}{2\sqrt{\delta E^{2}+\lambda^{2}/\pi^2}}\;.
\label{eq:phi}
\end{equation}
We then write the wavefunction as
$$\psi(\theta)=\frac{1}{\sqrt{2\pi}}\sin(\varphi/2)e^{in\theta}+\frac{1}{\sqrt{2\pi}}\cos(\varphi/2)e^{i(n+1)\theta}\;,$$
where $\varphi$ depends on $\lambda$ and $\Omega$.

The momentum distribution in this case becomes
\begin{eqnarray}
  \nonumber n(\textbf{k})&\sim &\left|\int_{0}^{2\pi}\mbox{d}\theta\;e^{i(k_{x}R\cos\theta+k_{y}R\sin\theta)}\psi^{*}(\theta) \right|^2\\
  \nonumber &=& \Big|\sin(\varphi/2)e^{in\gamma}J_{n}(|\textbf{k}|R)+\cos(\varphi/2)e^{i(n+1)\gamma}J_{n+1}(|\textbf{k}|R) \Big|^2\\
  \nonumber &=& \sin^{2}(\varphi/2)J^{2}_{n}(|\textbf{k}|R)+\cos^2(\varphi/2)J^{2}_{n+1}(|\textbf{k}|R)\\
  \nonumber &&+2\sin(\varphi/2)\cos(\varphi/2)\cos(\gamma)J_{n}(|\textbf{k}|R)J_{n+1}(|\textbf{k}|R)\;,\\
  \label{eq:nk}
\end{eqnarray}
where an interference term, proportional to $\cos\gamma$, appears between
the two states with defined angular momentum, giving rise to a $2\pi$-periodic
angular modulation of the ring shape found previously.
This behavior is the same found in Fig.~\ref{tofwithl}, where we observe
an analogous modulation in the weak barrier and weak interaction case,
that we can interpret than as direct consequence of the superposition
of two stated that differ by one quantum of angular momentum.

\section{Methods}
\label{methods}

\subsection{Infinite interaction limit: Tonks-Girardeau gas}

In the limiting case of infinite repulsive contact interaction between the particles
($U\rightarrow\infty$), the so-called hard-core bosons or Tonks-Girardeau gas,
an exact approach can be pursed to diagonalize Hamiltonian~\eqref{Model}.
Since multi-occupancy of one site is forbidden by the infinite interaction energy,
it can be simplified into
\begin{equation}
  H=-t\sum_{i=1}^{M}(e^{-i\Omega/M}b_{i}^{\dagger}b_{i+1}+h.c.)+t\sum_{i=1}^{M}\Lambda_{i} n^{b}_{i}
  \label{HCB}
\end{equation}
where the bosonic annihilation and creation operators have the additional
on-site constraints $b_{i}^{2}=b_{i}^{\dagger 2}=0$ and $b_{i}b_{i}^{\dagger}+b_{i}^{\dagger}b_{i}=1$.
By applying the Jordan-Wigner transformation
$$
b_{j}=\Pi_{l=1}^{j-1}e^{i\pi f^{\dagger}_{l}f_{l}}f_{j}\;,
$$
where $f_{i}$ ($f_{i}^{\dagger}$) are fermionic annihilation (creation) operators,
the Hamiltonian~\eqref{HCB} can be mapped into the one for spineless fermions:
\begin{equation}
  H=-t\sum_{i=1}^{M}(e^{-i\Omega/M}f_{i}^{\dagger}f_{i+1}+h.c.)+t\sum_{i=1}^{M}\Lambda_{i} n_{i}^{f}
  \label{F}
\end{equation}
This Bose-Fermi mapping is the analogous, for a discrete system,
of the one introduced by Girardeau for a continuous system~\cite{girardeau}.
Hamiltonians~\eqref{HCB} and~\eqref{F} have the same spectrum, but non-trivial differences
appear in the off-diagonal correlation functions: $\langle f^{\dagger}_i f_j \rangle$
vs $\langle b^{\dagger}_i b_j \rangle$, which we have calculated following
the same procedure described in~\cite{rigol}. Such difference affects,
for example, the momentum distribution, which is much narrower for hard-core
bosonic systems than for non-interacting fermions.
The density, and all the quantities related to it, are instead identical,
see for example Fig.~\ref{dens}. This 1D peculiar strongly correlated TG phase
has been demonstrated in several experiments on bosonic wires~\cite{paredes, kinoshita}.

\subsection{Gross-Pitaevskii equation}

In the limit of weak interactions, we adopt a mean-field approximation
to simplify the many-body Schr\"odinger equation.
This is the Gross-Pitaevskii (GP) equation for the bosons subjected
to a lattice potential, in the presence of a gauge field:
\begin{eqnarray}
  \frac{\hbar^2}{2m} \left(-i\partial_x-\frac{\Omega}{L}\right)^2\Psi +U_{b}\delta(x) \Psi + V_{0}\sin^{2}\left(\frac{M}{L}\pi x\right)\Psi+ g_{1D}|\Psi|^2 \Psi = \mu\Psi\;,
  \label{eq:gp1}
\end{eqnarray}
where $\Psi$ is the condensate wavefunction, $\mu$ is the chemical potential,
$V_0$ is the optical-lattice depth, $U_b$ is the strength of the localized barrier,
modeled as $\delta(x)$ in this continuous model, $m$ is the particle mass,
and $g_{1D}$ is the effective interaction coupling strength in one dimension,
related to the three-dimensional scattering length $a$ as $g_{1D}=2 \hbar^2 a/m a_\perp^2$~\cite{olshanii}.

The continuous-model barrier strength $U_b$ is connected to the discrete-model
one $\Lambda$ by $U_b=\Lambda L/M$. In absence of the lattice potential,
an analytical soliton solution for Eq.~\eqref{eq:gp1} has been found in~\cite{marco}.
In the further limiting case of vanishing interaction and small barrier strength,
the expression for the wavefunction can be obtained perturbatively with respect
to the barrier strength. This approach helps the understanding of the corresponding
momentum distribution and time-of-flight images (see~\ref{tof-nonint-l}).
In the presence of the lattice potential, we solve Eq.~\eqref{eq:gp1}
numerically by integrating it in imaginary times.
We pursue this approach as a benchmark case for the BH model at weak interaction.
Moreover the GP equation is a particularly suitable tool for the large-$N$ regime,
which is routinely realized in experiments.

\subsection{Exact diagonalization schemes}
\label{exact-diag}

\subsubsection{Working in the full Hilbert space}

The exact diagonalization (ED) is a computational method in many-body physics~\cite{Noack,Fehske}
which gives exact eigenstates and eigenvalues of the Hamiltonian without making
any simplifying assumptions about the physical system.
However the method is applicable to small systems and small fillings $N/M$.
The reason for that is provided by the fact that the Hilbert space spanned by
the many-particle Fock states cannot be too large. Specifically, to implement
the exact diagonalization, one has to consider all the possible combinations
of $N$ particles over $M$ sites, modulo the permutations of identical particles.
The dimension of the Hilbert space is given by~\cite{Zhang}:
\begin{equation}
D=\frac{(M+N-1)!}{(M-1)!N!}
\label{dimension}
\end{equation}
In Section~\ref{Sect:Momentum}, we considered values of the filling: $5/11, 15/11, 24/9$.
Correspondingly, the Hilbert space dimensions for that fillings are $3003; 3268760; 10518300$.

The non-diagonal part of the Bose-Hubbard model can be written efficiently
with the help of sparse matrices routine.
The ground and the first excited state state eigenvalues and eigenvectors
of the system can be found explicitly with help of Lanczos algorithm~\cite{Noack,Fehske}.

\subsubsection{Working in the truncated Hilbert space}

To study the system for larger sizes and larger fillings, the exact diagonalization scheme works
upon reducing the dimension of the Hilbert space in a controlled way.
This is achieved by restricting (truncating) the lattice site occupation number
up to some given integer number $K$.
The main difficulty of the truncated ED is the generation of the truncated
Hilbert space in a numerically efficient way~\footnote{One algorithm was suggested in A. Szabados, P. Jeszenski and P. Surjan,  Chem. Phys. {\bf 401}, 208 (2012).  It turns out, however, that that method is not efficient for generation of the big truncated Hilbert spaces ($D_K\sim 10^6$).}.
Here we are using the following algorithm to achieve the goal. At first we write
the function $f(M,n,K)$ which splits a positive integer $M$ into a sum of $n$ positive
integers (where each of the integers is smaller than $K$) up to commutativity
(so this function is returning matrix). Then we define the number $s$,
where $s=[M/K]+1$, if $M/K-[M/K]>0$ and else $s=[M/K]$.
Then the truncated Hilbert space can be generated in the following $3$ steps:
{\it 1)} to apply function $f(M,n,K)$ by changing $n$ from $s$ to $N$ with a step $1$;
{\it 2)} to concatenate each line in the matrix which return $f(M,n,K)$ with required
amount of zeros to make lines of matrix N dimensional arrays;
{\it 3)} to generate all possible permutations for any line of the matrix.
The dimension of the truncated Hilbert space is given by the following expression~\cite{Lew}:
\begin{equation}
  D_K=\sum_{i=1}^{\left[\frac{N}{K+1}\right]}(-1)^{j}\binom{M+N-1-j(K+1)}{M-1}\binom{M}{j}
  \label{dim1}
\end{equation}
where the brackets $[ ]$ stand for the floor function.

For example for the case of the filling $24/9$ and $K=6$ which is considered
in Section~\ref{Sect:Momentum}, $D_6=2345553$ which is almost $4.5$ times smaller
the dimension of the full Hilbert space. Indeed, in this way one can reduce $D$
for the several order of magnitudes, but that will introduce errors, especially at small $U$.
Here we estimate the errors in the following way. We calculate the particle number
fluctuations (variance) per lattice site
\begin{equation}
  \sigma_i=\sqrt{\langle n_i{^2} \rangle-\langle n_i \rangle ^2}
  \label{fluct}
\end{equation}
We assume that: if $\langle n_i \rangle+5 \sigma_i<K, \forall i$ then error
(in calculating expectation values) is smaller than $0.0006\%$,
if $\langle n_i \rangle+4 \sigma_i<K$ then error is smaller than $0.006\%$, if
$\langle n_i \rangle+3\sigma_i<K$ then error is smaller than $3\%$ and if
$\langle n_i \rangle+2\sigma_i<K$ for any i then error is smaller than $5\%$.
For the filling $24/9$ an estimated error for $U=10$ is less than $0.006\%$
and for $U=1$ it becomes $5\%$.

The momentum distribution in Fig.~\ref{tof_nu} is calculated with the scheme
detailed above.

\subsection{DMRG method}
\label{dmrg}

The modified Bose-Hubbard model of Eq.~\eqref{Model} can be quite naturally numerically
treated by a Density Matrix Renormalization Group (DMRG) approach, i.e.,
by optimising a Matrix Product State (MPS) representation of the many-body
wavefunction~\cite{Schollwock,Murg}.
A first requirement, as for almost any numerical treatment of bosons,
is to truncate the local Hilbert space down to few states, $d = n_{\max} + 1$,
with $n_{\max}$ the maximum allowed particle occupation per site.
The MPS ansatz is, at first sight, well suited for periodic boundary conditions~\cite{Verstraete}
but a practical implementation of an algorithm over it has to face a number
of subtleties and numerical instabilities~\cite{Murg}, especially in the case
of a non-homogeneous system, like here.
At a difference to another recent work of ours~\cite{marco}, then,
we decided to opt here not for an explicitly periodic MPS but rather to employ
a more standard open boundary (OBC) scheme with a trick.
Namely, we represented a ring of $M$ sites as two adjacent stripes
of length $M/2$ linked by only two extremal rungs on first and last site.

On one hand, such an approach implies an effective local dimension $d_\mathrm{eff} = d^2$,
as well as a larger bond dimension $m$ (i.e., the matrix dimension), and therefore
a priori an extra cost in the tensor contractions involved in the optimization process.
On the other hand, though, it allows us to rely on a unitary gauge from the left
and the right border, with deep computational advantages:\\
i) The optimization problem can be casted into a standard eigenvalue problem,
by means of exact contractions scaling as $O(m^3)$.
This has to be confronted with the situation for explicit PBC's:
there one has to face the instabilities of a generalised eigenvalue solver,
whose defining operators are moreover obtained exactly with $O(m^5)$ operations~\cite{Verstraete};
approximate, slightly better scaling, strategies for PBC's are also available,
but they work best on very long and uniform chains, which is not the case
here~\cite{Pippan,weyerauch}.\\
ii) The preservation of quantum numbers related to Abelian symmetries,
such as the $U(1)$ particle number, is simple to implement,
boost the computational efficiency~\cite{Schollwock}, and eliminates
an uncertainty source by not invoking any chemical potential~\cite{marco}.\\
iii) The splitting of a two-site (four-legged) optimised tensor into
two single-site (three-legged) ones assumes the clear meaning of an entanglement
renormalization (from $m \, d_\mathrm{eff}$ to $m$ states), thus giving a quantitative
indication of the performed approximation and permitting a dynamical allocation
of symmetry sectors inside the tensors~\cite{Schollwock}.
Such features compensate well the extra $d_\mathrm{eff}$ cost involved in the contractions
with respect to single-site optimization.

In our simulations~\footnote{The data presented in this work have been obtained by an open-source code available at \url{www.dmrg.it}}, we chose the local Hilbert dimension up to $d_\mathrm{eff} = 25$
(i.e., $n_{\max} = 4$) in the softer core case $U=1$:
this can be checked ``ex post'' to be appropriate, by looking at the decay of site
occupation probability and confronting it to other approximations incurring in the algorithm.
The other main source of numerical uncertainty is given, of course, by the number
of states kept in the RG procedure (i.e., the bond dimension of the MPS ansatz):
for moderate ring sizes up to 80-100 sites, as considered here, we have seen
that $m \simeq 200$ already provides reliable results.

\section*{References}
\bibliographystyle{iopart-num}
\bibliography{Paper}

\end{document}